\documentclass[10pt]{iopart}

\usepackage{cite}
\usepackage{url}
\usepackage{verbatim}
\usepackage{graphicx}
\usepackage{multirow}
\usepackage{makecell}
\usepackage{listings}
\usepackage{rotating}

\usepackage[caption=false,font=normalsize,labelfont=sf,textfont=sf]{subfig}
\usepackage{textcomp}
\usepackage{stfloats}
\usepackage[switch]{lineno}

\usepackage{xr}

\makeatletter
\newcommand*{\addFileDependency}[1]{
\typeout{(#1)}
\@addtofilelist{#1}
\IfFileExists{#1}{}{\typeout{No file #1.}}
}\makeatother

\newcommand*{\myexternaldocument}[1]{%
\externaldocument{#1}%
\addFileDependency{#1.tex}%
\addFileDependency{#1.aux}%
}

\myexternaldocument{icapipelines_JN_Supplementary}

\begin{document}

\title{I See Artifacts: ICA-based EEG Artifact Removal Does Not Improve Deep Network Decoding Across Three BCI Tasks}

\author{Taeho Kang\textsuperscript{1}, Yiyu Chen\textsuperscript{2}, Christian Wallraven\textsuperscript{2,3}*}

\address{\textsuperscript{1} Institute of Management Science, Technical University Wien, Vienna, Austria\\
\textsuperscript{2}
Department of Department of Artificial Intelligence, Korea University, Seoul, South Korea\\
\textsuperscript{3}
Department of Brain and Cognitive Engineering,  Korea University, Seoul, South Korea\\}
\ead{wallraven@korea.ac.kr}
\vspace{10pt}

\begin{abstract}

\textit{Objective.}
In this paper, we conduct a detailed investigation on the effect of IC-based noise rejection methods in neural network classifier-based decoding of electroencephalography (EEG) data in different task datasets.
\textit{Approach.}
We apply a pipeline matrix of two popular different Independent Component (IC) decomposition methods (Infomax, AMICA) with three different component rejection strategies (none, ICLabel, and MARA) on three different EEG datasets (Motor imagery, long-term memory formation, and visual memory). We cross-validate processed data from each pipeline with three architectures commonly used for EEG classification (two convolutional neural networks (CNN) and one long short term memory (LSTM) based model. We compare decoding performances on within-participant and within-dataset levels. 
\textit{Main Results.}
Our results show that the benefit from using IC-based noise rejection for decoding analyses is at best minor, as component-rejected data did not show consistently better performance than data without rejections---especially given the significant computational resources required for ICA computations.
\textit{Significance.}
With ever growing emphasis on transparency and reproducibility, as well as the obvious benefits arising from streamlined processing of large-scale datasets, there has been an increased interest in automated methods for pre-processing EEG data. One prominent part of such pre-processing pipelines consists of identifying and potentially removing artifacts arising from extraneous sources. This is typically done via Independent Component (IC) based correction for which numerous methods have been proposed, differing not only in the decomposition of the raw data into ICs, but also in how they reject the computed ICs. While the benefits of these methods are well established in univariate statistical analyses, it is unclear whether they help in multivariate scenarios, and specifically in neural network based decoding studies. As computational costs for pre-processing large-scale datasets are considerable, it is important to consider whether the tradeoff between model performance and available resources is worth the effort.

\end{abstract}

\noindent{\it Note\/}: This preprint is an outdated version from the final revision, as submitted by the author to Journal of Neuroal Engineering. IOP Publishing Ltd is not responsible for any errors or omissions in this version of the manuscript or any version derived from it. \textbf{The Recent Version of Record is available online at \url{https://doi.org/10.1088/1741-2552/ad788e}}. Please cite the article from this DOI link instead of the preprint.

%
%
%
%
\ioptwocol
%

\section{Introduction}

While many specific methods for pre-processing and artifact-removal of EEG data have been proposed, and guidelines for applying these methods do exist~\cite{keil2014committee, ortiz2013brain, gross2013good, kim2018preprocessing, bigdely2015prep}, the practical process and specifics of applying these methods have been left to the discretion of individual researchers, either through trial and error or by learning from examples made by their peers. It is an understandably challenging problem; new methods are created and published regularly, and it is a huge undertaking to compare and measure the effects different pre-processing methods have on the analysis result. Considering the configurable parameters that come with each method combined with the possible permutations of pre-processing methods chain one could employ in a pipeline, it is no wonder that creating a thorough, quantitative comparison between the different ensemble orchestrations of these methods so far has not been undertaken. To make matters worse, there exists no one true authoritative pre-processing pipeline, and it makes the already daunting task of comparing and replicating existing EEG study results even more difficult, as one must first figure out the specific pre-processing steps authors have employed. Several studies have attempted to tackle these issues to an extent by comparing the effects of several pre-processing methods on select datasets~\cite{bigdely2020automated, cashero2011comparison, murugappan2009comparison, wu2019explore, hoffmann2008correction, jung1998removing}. More recently, concerted efforts have been made to create a streamlined, automated and standardized process for data cleaning to pave way for a more reproducible research environment ~\cite{gabard2018harvard, da2018automatic, pedroni2019automagic, bigdely2015prep, winkler2014robust, pion2019iclabel}. 

In general, pre-processing in EEG is performed with the intent of improving the signal to noise ratio (SNR) of the data for analysis. This is especially common practice for analyses involving univariate statistical testing, such as event-related potential analyses of signals from different conditions. Scalp electrode measurements can be considered a mixture of signals whose sources are at a relative distance due to the structures between the cortex and the skin, and are much more susceptible to noise from movements such as ocular, muscular and cardiac activity than invasive measurements. It is also often subject to external noise such as those from electrical systems nearby (often referred to as power line noise), spurious physical contact of an object with electrodes, or even slow neurophysical activities and drifts~\cite{huigen2002investigation} that are usually not of interest. Filtering is one of the more common ways to deal with a number of different sources of noise, as well as enhancing features of brain signatures more apparent in certain frequencies depending on the topic of investigation. Notch filtering is common for removing powerline noise and its harmonics~\cite{ferdjallah1994adaptive}, and high pass filtering at a frequency around 1Hz is also common for removing slow components drifts~\cite{de2019filters}. For artifact removal, one of the simpler ways of removing noise is to visually inspect epochs and exclude channels and trials that are deemed too noisy. A similar outcome of labeling noisy epochs and channels can also  be achieved via automatic methods such as thresholding~\cite{krishnaveni2006removal}, although exclusion of data usually leads to loss of information and reduced amount of for analyses~\cite{nolan2010faster}. Blind source separation based methods alleviate this issue to an extent, via unmixing the sensor data into independent signals of estimated sources.
Artifact removal algorithms based on initial user annotations~\cite{somers2018generic} have also been found to improve signal quality with minimal distortions.
For a deeper look into filtering for EEG and biomodical signals, see Zhang et al.~\cite{zhang2024optimal}, as well as Cheveigne and Nelken ~\cite{de2019filters}. For a general introduction into removal of artifacts in EEG, see ~\cite{uriguen2015eeg}

Meanwhile, in {\emph decoding and classification} analysis studies, the benefits gained from various pre-processing methods are not be as clear. Findings from conventional EEG univariate statistical analysis methods are not always consistent with results from multivariate decoding analysis methods; as the latter often considers nonlinear effects on top of the linear, activation based data relationships of the univariate methods, information that may be considered as noise in univariate methods may not be so in decoding and classification tasks~\cite{davidson1972univariate, hebart2018deconstructing, gorgen2018same}. Such notion raises the question of whether the complex data cleaning and pre-processing procedures employed in classical brain signal analysis are actually also beneficial in classification studies.
In this context, Meisler et al. investigated the effect of initial pre-processing steps on classification performance: data rereferencing (bipolar and common average), noisy epoch exclusion (manual and statistics based automatic), channel exclusion (manual and automatic) on classification of a free recall task trained with logistic regression with intracranial EEG data (iEEG) - interestingly, results were varied and found to yield only minor improvements in classifier performance from channel or epoch exclusions~\cite{meisler2019does}.

Independent Component Analysis (ICA) is another method of reducing noise, which works by separating signals from multichannel data into time courses that are maximally independent from one another -- it assumes that by decomposing the signals it becomes possible to isolate the noise artifacts from brain signals (for a more technical overview of ICA in EEG analysis and BCI systems, see ~\cite{kachenoura2007ica, delorme2007comparing}).
ICA computation has long been used in EEG data analysis for noisy artifact detection and removal in EEG signals~\cite{iriarte2003independent, vigario1997extraction, delorme2007enhanced, vigario2000independent, castellanos2006recovering, makeig1996independent} and is one of the core functionalities offered in several scripting-based EEG analysis software packages~\cite{delorme2004eeglab, oostenveld2011fieldtrip, gramfort2014mne}. In addition, there are multiple implementations of different ICA-type algorithms available in these libraries alone: this includes Infomax~\cite{bell1995information}, extended Infomax~\cite{lee2000unifying} (both Infomax algorithms form the implementation runICA in EEGLAB), FastICA~\cite{hyvarinen2000independent}, AMUSE~\cite{cichocki2002adaptive}, ERICA~\cite{cruces2002robust}, and AMICA~\cite{palmer2012amica}.  
To illustrate the complexities inherent in the pre-processing pipeline, and also considering the significant computational cost for computing independent components in large EEG data compared to other methods, the present work focuses on the most common methods of ICA for artifact removal. 

Given the large variety of ICA algorithms alone, one may ask if and how the varying pre-processing pipelines for the same purpose of cleaning data affect the subsequent analysis results. To investigate this question, Bigdely et al. ~\cite{bigdely2020automated} used fully automated pre-processing pipelines and observed changes in channel amplitude dispersion from ICA-based ocular artifact rejections. Robbins et al.~\cite{robbins2020sensitive} studied the different data features and EEG signal characteristics yielded by several different data pre-processing methods (LARG~\cite{bigdely2020automated} and MARA~\cite{winkler2011automatic}) and two artifact subspace reconstruction-based (ASR)~\cite{kothe2016artifact} artifact removal methods on a total of 17 different datasets. Their findings show that the general characteristics of results from different pipelines of the same dataset were quite similar (though significantly different from the baseline of skipping the pipeline), reflected in significant correlations between ERP values; they also noted that LARG removed relevant EEG components as well while the other methods did not remove enough blink-related artifacts. Delorme's recent work reported a lack of improvements in the number of statistically significant ERP channels after automated ICA rejections were performed ~\cite{delorme2023eeg}. As demonstrated from these studies, not only does the application of a pre-processing step affect the resulting analysis, but different pre-processing pipelines serving the same purpose introduce some variance into the specifics of the resultant data.




In studies with classical machine learning methods, utilization of pre-processing methods that makes use of ICA computation has been reported to improve classifier performance.
For example, Kim et al. have reported that the removal of ocular artifacts using ICA leads to increased classification accuracy in MI tasks (and leading to more improvements upon combining the method with specific filters)~\cite{kim2017removal}. Another study by Kim and Kim has reported noise component removal on P300 oddball data leading not only to decreased variability in P300 amplitudes, but also to increased Brain-Computer Interface (BCI) classification accuracy when trained with support vector machines (SVM)~\cite{kim2018comparsion}. Winkler et al. observed pre-filtering the data with high-pass filter at low frequencies and applying ICA-based noise rejection afterwards with MARA resulted in improved signal-to-noise ratio (SNR) and linear discriminant analysis (LDA) classifier performance on auditory oddball task data~\cite{winkler2015influence}.

Recently there has been an influx of studies utilizing deep neural networks for brain state classifications (see ~\cite{craik2019deep, altaheri2021deep} for review). 
Here, the question of whether applying pre-processing methods result in better analysis, or classifier performance, is becoming more frequently raised, and the results are not as clear as in classical machine learning.
In an end-to-end deep learning study with data from a Stroop task, Fahimi et al. reported pre-processing EEG data by applying narrower pass band filters resulted in decreased decoding accuracy levels in convolutional neural networks (CNNs)~\cite{fahimi2018inter}.
In a review of deep learning classification methods for Motor Imagery (MI) signals, Altaheri et al. discusses the possible effects of several pre-processing methods on network classifier performance in MI data~\cite{altaheri2021deep}: Xu et al.~\cite{xu2020learning} reported filtering for narrower frequency ranges results in better CNN performance in MI, whereas studies such as Riyad et al.~\cite{riyad2021mi}, Avilov et al.~\cite{avilov2021optimizing}, and Shajil et al.~\cite{shajil2020multiclass} report wider frequency bands yield higher classification accuracy.
Zhu et al.~\cite{zhu2019study} observed that having a large number of channels regardless of assumed relevance to spatial brain region of interest results in higher classifier performance in CNN and LSTM based models. Pfurtscheller et al.~\cite{pfurtscheller2006mu} reported reducing the number of input EEG channels resulted in higher classifier performance in some deep-belief network (DBN) models, while yielding reduced performance in other models of similar design. The same study also reported that artifact removal with synchrosqueezed wavelet transforms (SWT) resulted in improved decoding performance in some participants, but reduced performance in others, suggesting EEG artifact removal may need to be revisited for deep network model classifiers.

%
%
 Among the standard different pre-processing methods available for EEG data, ICA-computation is a computational and time resource intensive process~\cite{laparra2011iterative}, especially as the number of samples and participants increase. For efficient use of valuable computation time, especially in multi-dataset classification study use cases where univariate statistical testing of epoched data may not be necessary, knowing whether a pre-processing step is really necessary could save a significant amount of resource. 
 In light of the considerations for decoding studies, as well as the opposing findings on the benefit of noise correction methods between studies performed with classical machine learning algorithms and those performed with deep networks, the goal of the present study was to investigate whether ICA-based noise correction methods do result in improved deep network model classification performance for BCI tasks.
The objective of this study is twofold: First, to investigate the effect of automated ICA-component rejection based EEG noise cleaning methods on classifier performance - at the same time we take care to minimize the variance in other pre-processing methods between datasets as well as to keep explicit feature extraction to a minimum by applying an otherwise standardized automated pre-processing pipeline.  Second, to investigate whether the effect of such methods are generalizable across different datasets and different network architectures. 

 Given the findings in prior deep networks studies where mixed results were reported regarding the benefit of pre-processing methods, we hypothesized that ICA-based noise rejection methods may not necessarily benefit classification performance in deep networks. We tested this notion on three different BCI task databases; first, a dataset from the BCI-competition IV-2a (abbreviated as BCIC-IV2a) dataset for Motor Imagery (MI) that is a standard benchmark dataset for EEG decoding tasks, and second, a long-term memory prediction dataset that we created in a separate paper~\cite{kang2020eeg}, and third, a visual memory prediction dataset that is to be published from a study comparing the effect of different EEG pipelines on analysis results~\cite{trubutschek2022eegmanypipelines}. For a comparison between different ICA noise correction strategies (and to minimize human bias in the process of noise correction), we used a combination of two different methods of independent component (IC) computation (RunICA of InfoMax, and AMICA), and two different automated IC based noise rejection methods (MARA, ICLabel). Combined with the baseline pipeline in which no ICA-based noise correction was performed, we created a total of 5 different ICA-noise pipelines for each dataset. And to test whether the resultant effect on network performance was not architecture dependent, we trained and optimized hyperparameters for each participant of each dataset for each ICA pipeline on 3 different network designs: ShallowNet, MLSTM-FCN, and EEGNetV4. We performed analyses on classifier performance metrics on participant individual level across pipelines, as well as on a per-dataset level. With such automated methods, we designed the study to have minimal human experimenter input in the actual noise rejection, in order to not only minimize human bias, but also in the interest of keeping the study reproducible.

\subsection{Independent Component Analysis}
Blind source separation refers to the process of which individual sources of signals are separated from measured data that contains mixtures of both noise and signals of interest.
Independent Component Analysis (ICA) is one such method used to perform this task, and is commonly applied in the processing of biomedical data~\cite{naik2014blind, ye2012heartbeat, li2011application} such as EEG signals~\cite{james2003temporally, jung2000removing, vazquez2012blind} where observed data usually involves multi-channel data that is a variable mix of multiple signal sources. The general idea of ICA is that signals are separated in an unsupervised manner, under the assumption that sources of original signals are statistically independent and have non-Gaussian distributions, and that each of these original and independent sources derive from different underlying physical processes. 
In EEG, observations of these signals come from spatially distributed sensors, each measuring a different mixture of source signals coming from supposedly separate neural processes. Here, the matrix containing observed signals from the different sensors can be described as a product of an unknown matrix (commonly termed the mixing matrix in this context) and the matrix of the original source signals. Assuming the unknown mixing matrix is invertible, and by estimating the inversion of the mixing matrix (dubbed the unmixing matrix), an approximation of the source signals can be calculated as a product of the measured data and the unmixing matrix. This is the general process in which approximated source signals are calculated through ICA. For a fundamental overview on ICA, see ~\cite{naik2011overview}. For an in-depth analysis of comparing different ICA methods on EEG signals, see~\cite{albera2012ica, delorme2012independent}.
While there are numerous ICA methods to achieve signal separation such as Infomax~\cite{bell1995information}, FastICA~\cite{hyvarinen2000independent}, SOBI~\cite{belouchrani1997blind}, ERICA~\cite{cruces2002robust}, JADE~\cite{cardoso1993blind}, and AMICA~\cite{palmer2012amica, palmer2007modeling}, in this paper we utilize implementations of two methods to compute independent components in EEG: RunICA (Infomax) and Adaptive Mixture ICA (AMICA). These two methods have been compared regarding their effect in terms of reduction of mutual information between components~\cite{delorme2012independent} and a measure of SNR~\cite{leutheuser2013comparison}. Specifically, in Delorme et al., of the multiple ICA algorithms compared for the amount of reduction in mutual information in EEG the two algorithms had the highest values ~\cite{delorme2012independent}. Leutheuser et al. reported a higher measure of SNR from data that were processed with AMICA instead of Infomax~\cite{leutheuser2013comparison}.

\subsection{Infomax}
The Infomax method of IC decomposition is based on the algorithm by Bell and Sejnowski~\cite{bell1995information}, in which the unmixing matrix is optimized via gradient ascent so that the joint entropy between approximated source signals is maximized with the goal of minimizing mutual information between measured data and estimated source.
RunICA is the implementation of Infomax ICA computation method provided by EEGLAB~\cite{delorme2004eeglab} that also optionally incorporates findings from Lee et al.~\cite{lee1999independent} and Amari et al.~\cite{amari1995new}. Amari that were et al.'instead of applies natural gradient descent, which is the standard gradient multiplied by the inverse of the Fisher information matrix, with the reported advantage that fewer iterations are needed compared to the traditional gradient method~\cite{amari1998natural, martens2020new}. Lee et al.'s extended Infomax method incorporates two learning rules to allow for signals with super-Gaussian and and sub-Gaussian distributions~\cite{lee1999independent}. 

\subsection{Adaptive Mixture ICA (AMICA)}
Adaptive Mixture ICA takes a probabilistic approach to compute independent component by the use of mixture models (a group of methods also known as ICAMM), with the advantage that frequency components in the source data are allowed to be dependent while at the same time preserving independence between sources~\cite{kim2006independent, shah2002ica, lee2000ica, palmer2007modeling}.
Specifically, AMICA utilizes Gaussian scale mixture (GSM)~\cite{portilla2003image} with the incorporation of stability analysis from Amari et al.~\cite{amari1997stability}, allowing to potentially construct multiple models for ICA and choose for more accurate source density models~\cite{palmer2007modeling}. Implementation is available in EEGLAB.

\subsection{Automatic EEG pre-processing and IC based noise rejection methods}
Over recent years, several methodological frameworks to introduce automation and standardization into the pre-processing steps of EEG data have been introduced. Different methods span varying portions of the EEG data handling process, with some incorporating most of the pre-processing steps that are commonly performed, while others are designed strictly for more specific pre-processing operations such as artifact rejection. Here we look at such examples: HAPPE, PREP, Faster, MARA, and ICLabel.
The Harvard Automated pre-processing Pipeline for EEG (HAPPE)~\cite{gabard2018harvard} is an integrated framework that incorporates different open-source EEG processing method implementations to provide a standardized pipeline for easier replication of studies. At the time of its original publication the framework supported processing of multi-channel EEG data with at least 64 channels, and the framework involves most of the pre-processing steps in EEG; filtering, channel selection, line-noise removal, channel rejection, ICA and component rejection through MARA~\cite{winkler2014robust}.
The PREP pipeline~\cite{bigdely2015prep} provides a basic and standardized framework for earlier stages of EEG pre-processing such as line-noise removal, detrending, re-referencing, and channel interpolation for noisy data.
Fully Automated Statistical Thresholding for EEG artifact Rejection (FASTER) is an automated framework that focuses on noise removal from data during pre-processing based on statistical estimations of data characteristics~\cite{nolan2010faster}. At different levels of the data (channels, epochs, independent components, channels in epochs, and grand average), statistical parameters of the data such as variance, mean correlation, Hurst exponent are calculated. Based on the thresholds of z-scores for each computed parameter, data with noise are identified and removed, interpolated, or subtracted depending on the level of the data under inspection.
Both MARA and ICLabel are classifier-based methods for independent component identification and rejection of noise in EEG. The former, Multiple Artifact Rejection Algorithm~\cite{winkler2011automatic, winkler2014robust} computes a set of features from the computed independent components, such as current density norm, logarithmic range within pattern, mean local skewness, logarithmic alpha band power, and fit error. A supervised linear binary classifier pre-trained on aforementioned features of 840 expert-labeled EEG independent components is used on said features of input data to determine noise components. ICLabel~\cite{pion2019iclabel} is another classifier-based noisy independent component rejection method that utilizes artificial neural networks. The classifier is trained on over 200,000 independent components extracted from 6300+ recordings of EEG from various channel number setups between 32 and 256. Inference is made by assigning probability values between 7 classes for each component: Brain, Muscle, Eye, Heart, Line-noise, Channel-Noise, and Other. Components with classes other than Brain are considered noise signals, and the user may set probability threshold values for noise classes to reject components post-inference. 

In this paper, we make use of the PREP pipeline to process data through the initial pre-processing steps, then compute independent components through two methods: AMICA and RunICA (Infomax). Finally, we reject components by the two classifier based methods MARA and ICLabel. We applied PREP beforehand in order to keep pre-processing steps before ICA across datasets as consistent as possible, with minimum manual input, chose the two ICA computation methods based on existing ICA method studies which used them~\cite{delorme2012independent, leutheuser2013comparison}. We also selected the two automated ICA-based rejection methods after considering the number of citations on the papers, the download counts in EEGLAB's add-on feature menu, and their usage in published automated pre-processing pipeline studies~\cite{gabard2018harvard, rodrigues2021epos}.

\section{Methods and Materials}

\begin{figure*}
	\centering
	\captionsetup{justification=centering}
		\includegraphics[width=0.8\textwidth]{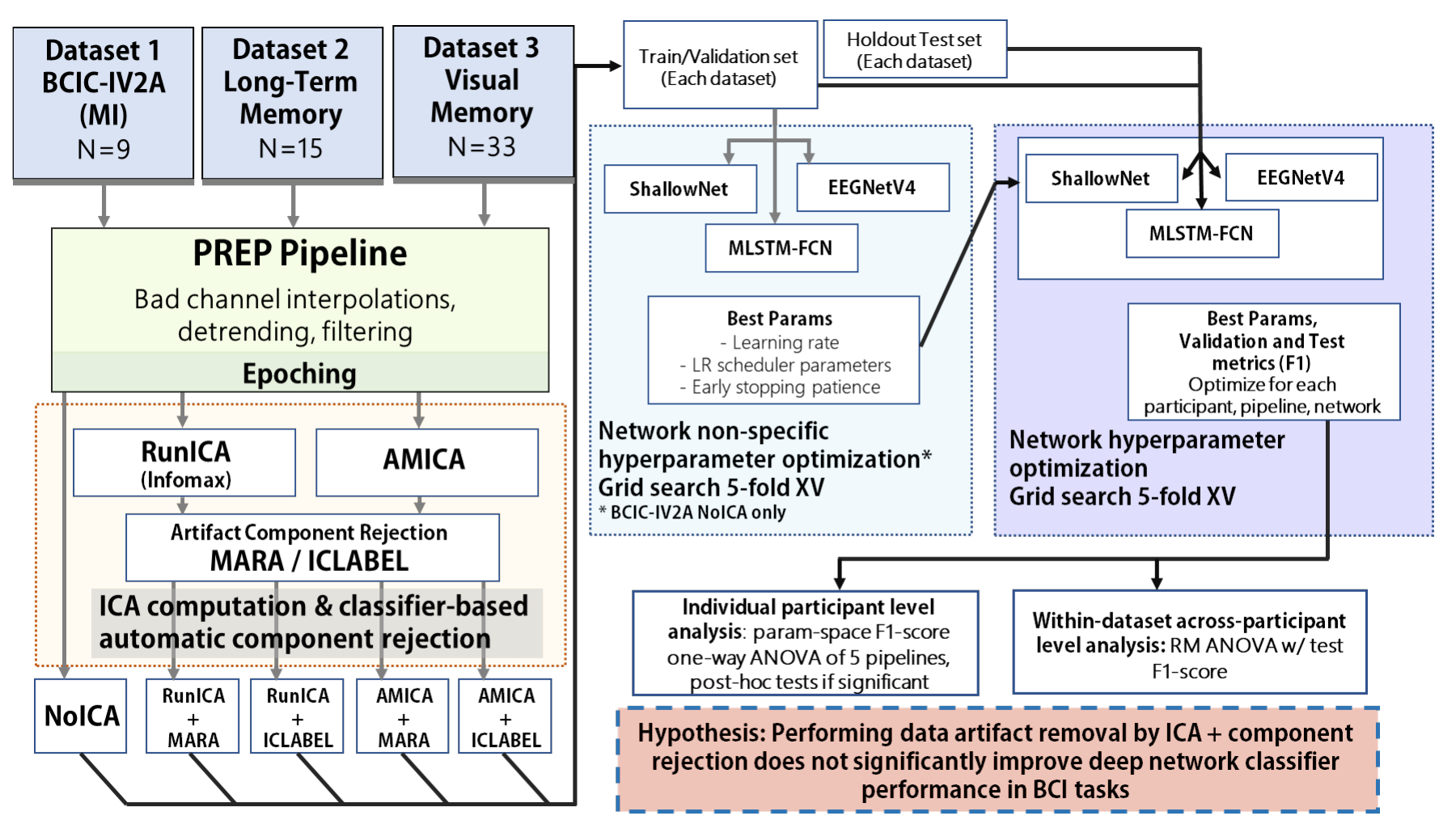}	
	\caption{Study overview}
	\label{fig:study_overview}
\end{figure*}

\subsection{Used datasets, data pre-processing pre-ICA component rejection pipelines}
To compare the effect of ICA noise rejection pipelines on EEG datasets underlying different tasks and brain dynamics as well as to investigate the possibility of a generalized result, the BCI competition dataset IV 2a (Motor Imagery )\cite{brunner2008bci}, a long-term memory prediction dataset~\cite{kang2020eeg}, and a visual memory dataset from EEGManyPipelines~\cite{algermissen2021eegmanypipelines, trubutschek2022eegmanypipelines} were used. The BCI competition dataset IV 2a consists of 9 participants worth of data recorded with 22-EEG channels (+3 EOG), with one training session and one test session for classifier validation. A non-epoched format of the BCIC-IV2a dataset was available for us to apply the automated pre-processing steps appropriate for our purpose (note that this data is already bandpass-filtered, however, as the raw data is not available). The long-term memory dataset consisted of 15 participants (1 additional participant was added after the publication of the original dataset paper) with 62-channel EEG (+1 EOG) data, each with 15 sessions of experiment performed across 5 consecutive days. The visual memory dataset consisted of 64-channel EEG data from 33 participants who performed recognition tasks on 600 images of scenery. We selected these datasets to encompass a reasonable variety of tasks with different levels of complexity in involved neural processes without overwhelming our time and compute resources, with a preference for public availability, as well as considerable sample size which is strongly preferred in machine learning models.

First, data files from each dataset were pre-processed through the automated PREP pipeline from EEGLAB~\cite{bigdely2015prep} for a standardized pre-processing step using the default running option parameters. Specifically, detrending was performed on all EEG channels with a high-pass cutoff of 1Hz. Powerline noise and its harmonic frequencies were notch filtered, and based on robustly-referenced signals noisy channels were automatically removed and interpolated from the remainder. The BCIC-IV2a dataset was downsampled to 250Hz, the long-term memory dataset and the visual memory dataset were downsampled to 100Hz. Next, data files were epoched in relation to the markers of interest in the respective datasets: in the BCIC-IV2a dataset, time series epochs of -500ms pre-stimulus and 5000ms post-stimulus were created in relation to the stimulus onset. In the long-term memory dataset, feedback-locked epochs from -200ms to 1000ms in relation to feedback onset were used for epoching, as it was in line with the reference paper of the dataset~\cite{kang2020eeg}. For the visual memory dataset, stimulus onset locked epochs from -200ms to 1000ms were extracted.

Including a raw no-ICA pipeline where no component rejection was performed as control, a total of five ICA-related artifact removal pipelines were created, using combinations of two prominent ICA methods in EEG and two automatic classifier-based artifact component rejection methods: AMICA~\cite{palmer2012amica}, and the runICA implementation of Infomax ICA ~\cite{bell1995information} for ICA computation, and MARA~\cite{winkler2011automatic}, and ICLabel~\cite{pion2019iclabel} for automated classifier based component rejections. Independent component computations were performed independently for each EEG recording session worth of data. Automated component rejection methods were used to minimize human experimenter bias in noise, as well as in the interest of reproducibility of the study.
As ICLabel is a multi-class classifier that predicts probability values for several different types of artifact signals as well as the brain signal, we chose a probability threshold of $80$\% for artifacts as the rejection criteria. That is, if a given component's class probability for labels other than "brain" and "other" was above 80\%, it was marked as an artifact and rejected. It should be noted however, that there is no definitive way to reach the final decision using the ICLabel's results, such that our choice of threshold was based on best practices (see, for example, ~\cite{pernet2021bids, papin2024investigating, chatzichristos2020epileptic}, but contrast with \cite{thammasan2020cross}). The full summary of the study pipeline can be seen in Figure~\ref{fig:study_overview}. 

\subsection{Classifier training and comparison}
While a certain amount of pre-stimulus (and pre-feedback) portion of the data were included in the epoch for ICA computations and noise rejections, only the post-stimulus and post-feedback onset portion of the data were used in the classifier training segment of the study. 

In order to investigate whether any effect to classifier performance was observable across different neural network designs, we utilized three different network architectures separately for each dataset and pipeline: ShallowNet ~\cite{schirrmeister2017deep}, EEGNetV4 ~\cite{lawhern2018eegnet}, and MLSTM-FCN ~\cite{karim2019multivariate}. ShallowNet and EEGNetV4 were developed specifically for EEG signal classification purposes, whereas MLSTM-FCN was developed for time-series classification in general. For all networks, the Pytorch~\cite{paszke2019pytorch} implementation was used. Specifically, for EEGNetV4 and ShallowNet implementations, Pytorch-based instances from Braindecoding~\cite{schirrmeister2017deep} version 0.6 were used. 

To ensure observed results could be reproducible and comparable across different pipeline settings, we pre-configured several options in Pytorch as well as in Python: Before hyperparameter optimization code corresponding to each dataset's ICA pipeline ran, Numpy's randomness seed, as well as base Python's and Pytorch's randomness seed values were all locked into a specific value (4224). Furthermore, the backend determinism option of pytorch (torch.backends.cudnn.deterministic) was set to True. 
For each participant data, a holdout test-set was defined for final performance evaluation metric. In the BCIC-IV2a dataset, the test session data for each participant was used for this. In the long-term memory dataset, 10\% of the data were set aside as a hold-out test set using scikit-learn's Stratified Shuffle split function. The remaining data was split into training and validation sets using stratified split to be used for general hyperparameter optimization in a 5-fold cross validation grid search scheme.

To account for possible class imbalance in data, weighted cross entropy loss was weighted in regards to ratio of classes within the dataset:
\[
W_i = 1 - \frac{N_i}{\sum N_i}
\]
where $W_i$ denotes the weight for a given class $i$, and $N_i$ denotes the trial sample size of the class $i$.
No data augmentation schemes, or balancing by under/oversampling of either numerous or fewer class were used. All 3 networks were optimized with the AdamW optimizer~\cite{loshchilov2017decoupled}.

\begin{table}
	\scriptsize	
	\captionsetup{justification=centering}
	\centering
	\begin{center}
			\begin{tabular}{|l|c|c|c|}				
				\hline	
				\multicolumn{4}{|c|}{\textbf{All networks, NoICA BCICIV-2A only}}\\
				\hline
				Parameter&\multicolumn{3}{|c|}{Values}\\				
				\hline
				Initial learning rate&\multicolumn{3}{|c|}{.3, .1, .01, .005, .001}\\				
				\hline
				${cos}$-annealing LR scheduler parameter T0&\multicolumn{3}{|c|}{10, 20, 30}\\				
				\hline
				${cos}$-annealing LR scheduler parameter $T_m$&\multicolumn{3}{|c|}{2, 4, 6}\\				
				\hline
				Early stopping epoch patience&\multicolumn{3}{|c|}{20, 40, 120, 240}\\					
				\hline
				Total parameter space&\multicolumn{3}{|c|}{180}\\				
				\hline
			\end{tabular}
	\end{center}
	
	\caption{Network agnostic hyperparameter search} 
	\label{tab:lr_serach_space}	
\end{table}

\begin{table}
	\scriptsize	
	\captionsetup{justification=centering}
	\centering
	\begin{center}
			\begin{tabular}{|l|c|c|}				
				\hline	
				\multicolumn{3}{|c|}{\textbf{ShallowNet, all ICA pipelines}}\\
				\hline
				Parameter&\multicolumn{1}{|c|}{BCIC IV2A}&\multicolumn{1}{|c|}{Memory (both)}\\				
				\hline
				filter\_time\_length&\multicolumn{1}{|c|}{15, 25, 35}&\multicolumn{1}{|c|}{4, 8, 16}\\				
				\hline
				n\_filters\_spat&\multicolumn{1}{|c|}{20, 40}&\multicolumn{1}{|c|}{20, 40}\\				
				\hline
				n\_filters\_time&\multicolumn{1}{|c|}{20, 40}&\multicolumn{1}{|c|}{20, 40}\\				
				\hline
				pool\_time\_length&\multicolumn{1}{|c|}{20, 40, 75, 80}&\multicolumn{1}{|c|}{4, 8, 12}\\				
				\hline
				pool\_time\_stride&\multicolumn{1}{|c|}{5, 10, 15}&\multicolumn{1}{|c|}{2, 4, 8}\\				
				\hline
				Total parameter space&\multicolumn{1}{|c|}{144}&\multicolumn{1}{|c|}{108}\\				
				\hline
				\hline
				\multicolumn{3}{|c|}{\textbf{MLSTM-FCN, all ICA pipelines}}\\
				\hline			
				Parameter&\multicolumn{1}{|c|}{BCIC IV2A}&\multicolumn{1}{|c|}{Memory (both)}\\				
				\hline
				conv1\_size&\multicolumn{1}{|c|}{64, 128}&\multicolumn{1}{|c|}{64, 128}\\				
				\hline
				conv2\_size&\multicolumn{1}{|c|}{128, 256}&\multicolumn{1}{|c|}{128, 256}\\				
				\hline
				conv1\_kernel&\multicolumn{1}{|c|}{8, 24, 32}&\multicolumn{1}{|c|}{6, 12, 24}\\				
				\hline
				conv2\_kernel&\multicolumn{1}{|c|}{5, 15}&\multicolumn{1}{|c|}{4, 8}\\				
				\hline
				conv31\_kernel&\multicolumn{1}{|c|}{3, 9}&\multicolumn{1}{|c|}{3, 6}\\				
				\hline
				lstm\_hidden\_size&\multicolumn{1}{|c|}{8, 16, 24}&\multicolumn{1}{|c|}{8, 16, 24}\\				
				\hline
				lstm\_num\_layers&\multicolumn{1}{|c|}{1, 2}&\multicolumn{1}{|c|}{1, 2}\\				
				\hline
				Total parameter space&\multicolumn{1}{|c|}{288}&\multicolumn{1}{|c|}{288}\\				
				\hline
				\hline
				\multicolumn{3}{|c|}{\textbf{EEGNetv4, all ICA pipelines}}\\
				\hline			
				Parameter&\multicolumn{1}{|c|}{BCIC IV2A}&\multicolumn{1}{|c|}{Memory (both)}\\				
				\hline
				F1&\multicolumn{1}{|c|}{4, 8, 16}&\multicolumn{1}{|c|}{4, 8, 16}\\				
				\hline
				D&\multicolumn{1}{|c|}{2, 4, 8}&\multicolumn{1}{|c|}{2, 4, 8}\\				
				\hline
				F2&\multicolumn{1}{|c|}{8, 16, 24}&\multicolumn{1}{|c|}{8, 16, 24}\\				
				\hline
				kernel\_length&\multicolumn{1}{|c|}{16, 32, 64}&\multicolumn{1}{|c|}{8, 16, 24}\\				
				\hline
				third\_kernel\_size&\multicolumn{1}{|c|}{[2,4], [4,2], [8,4]}&\multicolumn{1}{|c|}{[2,4], [4,2], [8,4]}\\				
				\hline
				Total parameter space&\multicolumn{1}{|c|}{243}&\multicolumn{1}{|c|}{243}\\				
				\hline
				\hline
				\multicolumn{3}{|c|}{\textbf{Network/Data non-specific parameters, locked}}\\
				\hline			
				Parameter&\multicolumn{2}{|c|}{Values (Common)}\\				
				\hline
				Initial learning rate&\multicolumn{2}{|c|}{0.001}\\				
				\hline
				Cosine annealing T0&\multicolumn{2}{|c|}{10}\\				
				\hline
				Cosine annealing $T_m$&\multicolumn{2}{|c|}{6}\\				
				\hline
				Early stopping patience&\multicolumn{2}{|c|}{240}\\					
				\hline
			\end{tabular}
	\end{center}
	
	\caption{Network and dataset dependent hyperparameter space} 
	\label{tab:net_param_space}	
\end{table}

Hyperparameter optimization was performed in two stages: one preliminary stage for setting general network non-specific hyperparameters with a single pipeline of a single dataset (BCIC-IV2a, NoICA), and the main hyperparameter optimization stage, in which network-specific parameters were optimized for each network in each participant of datasets pre-processed through each ICA rejection pipeline. 

In the preliminary stage, we performed gridsearch to set common hyperparameters to be shared across the 3 different architectures: the initial learning rate, parameters $T_m, T_0$ for the cosine annealing LR scheduler with warm restarts~\cite{loshchilov2016sgdr}, and the patience epoch value for triggering early stopping when validation loss did not improve for a certain amount of epochs. Specific hyperparameter space values for the above can be found in Table~\ref{tab:lr_serach_space}. Gridsearch for these parameters were done in a separate preliminary level, trained with only one pipeline of a single dataset under the assumption that the optimal parameters for these would not vary as much as the network specific parameters, as well as to save on optimization time as combining these with network specific hyperparameter space would create too large of a search space. Only the training and validation sets were used in this stage, as the final performance comparison was to be performed after the main hyperparameter optimization process.

In the main hyperparameter optimization stage, we tuned each instance of individual networks for each dataset-participant and for each ICA-rejection pipeline separately. The specific hyperparameter search space for each network can be found in Table~\ref{tab:net_param_space}. Validation F1-score metrics were used to find the best performing hyperparameter set, and once the best performing model was chosen, the holdout-test sets were predicted for the final performance measurement (Test-set F1 score). The gridsearch procedure was implemented with Skorch~\cite{skorch}. In each parameter set, training was terminated early once there was no validation loss improvement for epochs equal to the set early stopping patience value. Neural network optimizations were performed on 6 NVIDIA 3090 and one 3090Ti machines. Across all participants and all pipelines, 360 iterations of gridsearch spanning a total search space of 78,300 parameter combinations were performed. Implementations and code for the general study pipeline can be found at \url{https://github.com/52bloo/eeg_icacomparison}.

\subsection{Statistical Analysis}
Statistical analyses of the main hyperparameter optimization result were performed in two stages: on individual dataset-participant level for the entire gridsearch space, and on across-participant level for each network using performance evaluation metrics of best-performing validation F1 and test-set F1 scores. On individual-participant-level analysis, one-way analysis of variance (ANOVA) was performed for each participant's results from a given network design comparing validation F1 scores from the entire gridsearch process between the 5 different ICA rejection pipelines: NoICA, RunICA+MARA, RunICA+ICLabel, AMICA+MARA, AMICA+ICLabel with an $\alpha$ threshold of 0.05. Post-hoc Tukey's HSD tests were administered for each participant to determine if there were specific pairs of pipelines that yielded significant difference in performance metrics. Correlation matrices were computed for the best-performing validation F1 score and test F1 score across pipelines and networks. On the cross-participant-level analysis, repeated measure analysis of variance (RM-ANOVA) was performed across pipelines for each network and dataset, using results from each participant as a repeated measure and pipelines as the group variable. Post-hoc tests on this level were performed with pairwise T-tests, in which the p-values were corrected for multiple comparison with Bonferroni correction. As we counted ANOVA runs from different networks and datasets as separate from multiple comparison, the Bonferroni correction multiplier was computed as the total number of comparisons across pipelines (combinations of 2 from 5 = 10). ANOVA and post-hoc Tukey tests used implementations from the statsmodel library ~\cite{seabold2010statsmodels}. 

\section{Results}

\subsection{Component Rejections} 

In the BCI Competition IV-2A dataset, an average of 0.1 artifact components were rejected from AMICA component data rejected with ICLabel, while an average of 4.9 artifacts components were rejected through MARA. In data processed with RunICA, an average of 0.1 components were rejected through ICLabel, and 5.1 through MARA. As the number of rejected components did not follow a normal distribution, non-parametric Mann Whitney U tests were performed to test for differences in the number of components rejected. A significant difference in the number of components rejected between MARA-based and ICLabel-based pipelines were found ($p<1e^{-13}, z=-7.660$), while there were no significant differences between data processed through the same rejection algorithms but different component calculation methods(RunICA, AMICA) ($p=.971, z=.0355$). Specifics for the component rejections can be found in Supplementary Materials Table 1.

In the AMICA-processed components of the long-term memory dataset, an average of 4.9 components were rejected with ICLabel, while an average of 36.6 components were rejected by MARA. A similar pattern of rejected component counts were observed in the RunICA pipeline; 5.2 components in ICLabel and 35.1 components in MARA. Similar patterns of difference in rejected component counts were also visible on Mann-Whitney U Test results: a significant difference between MARA-based and ICLabel-based rejections ($p<1e^{-149}, z=-25.976$), and a non-significant difference between RunICA and AMICA ($p=.719, z=.359$). Specifics for the component rejections can be found in Supplementary Materials Table 2,3,4, and 5.

For the visual memory dataset, an average of 5.7 components were rejected through ICLabel on components computed with AMICA, while an average of 33.4 components were rejected from the same component computations with MARA.
In component data processed with RunICA, an average of 5.8 components were rejected through ICLabel, while an average of 35.5 components were rejected through MARA. 
Again, Mann-Whitney U tests of the rejected component counts showed similar findings: there were observed significant differences between the component rejection methods ($log(p)<-23$, $z=-9.92$)  while no such observations were made in comparisons of values between ICA computation methods ($p>.5$, $z=-.52$). The detailed values of component rejections can be seen in Supplementary Materials Table 6.

\subsection{Preliminary network non-specific hyperparameter search}

Detailed results for the optimization of network non-specific parameters (learning rate, cosine-annealing learning rate scheduler parameters T0 and TM, early stopping patience epoch) can be found in the Supplementary Figures 1 (ShallowNet), Figure 2 (MSLTM-FCN), and Figure 3 (EEGNet). 
Overall, we found that the lowest candidate learning rate and the highest early stopping patience value were similar across all three networks. The learning rate scheduling parameters did not appear to noticeably impact the validation performance within participants. Considering the apparent advantage of lower learning rates and higher early stopping patience, and of introducing resetting of the learning rate early on in the epochs while gradually decreasing its frequency, we chose the following parameters for the main hyperparameter optimization procedure across all 3 networks: an initial learning rate of 0.001, Cosine annealing LR scheduler T0 of 10, $T_m$ of 6, and an early stopping patience value of 240 epochs.

\subsection{Main hyperparameter search: individual level comparison of classification accuracy across pipelines}
A detailed, participant-individual-level visualization of the main hyperparameter search is shown in Supplementary Figure 4(BCIC-IV2a), Supplementary Figure 5 (Long-term memory), and Supplementary Figure 6 (Visual memory) for each dataset. 
In all participants across all networks and pipelines in each dataset, holdout test-set performance was sufficiently above chance when compared to confidence-level upper limit of simulated chance levels using ~\cite{muller2008better}'s method. In the BCIC-IV2a dataset, one-way ANOVA performed on each participant's all validation F1 scores from the gridsearch space in each network across pipelines found significant p-values lower than 0.05 in 5 out of 9 participants in Shallownet ($p>.05$ in ptc. 1, 3, 6, 9), 7 out of 9 for MLSTM-FCN ($p>.05$ in ptc. 6, 9), and 7 out of 9 for EEGNet ($p>.05$ in ptc. 6, 9). Interestingly, ANOVA results were not significant in all 3 networks for participant 6 and 9. In the same analysis performed on the long-term memory dataset, 14 out of 15 participants had p values lower than 0.05 in EEGNet ($p>.05$ in ptc. 14) and Shallownet ($p>.05$ in ptc. 10), while all 15 participants had significant p-values in MLSTM-FCN. As can be seen in the circos plots, the chosen best performing parameters were highly variable between participants (and somewhat even across pipelines in same participant) for MLSTM-FCN and ShallowNet, while they were somewhat consistent for EEGNetV4 in both datasets. For EEGNet, the chosen parameters were more consistent across pipelines for the same participant as well. 
In results from the visual-memory dataset, ANOVA from every participant in each network showed significant p-values ($p<.05$) from the F1 scores in the gridsearch space.
Further look into the best performing hyperparameters can be found in discussions.

\begin{figure*}[htbp]	
	\centering
	\captionsetup{justification=centering}
		a)\includegraphics[width=0.3\textwidth]{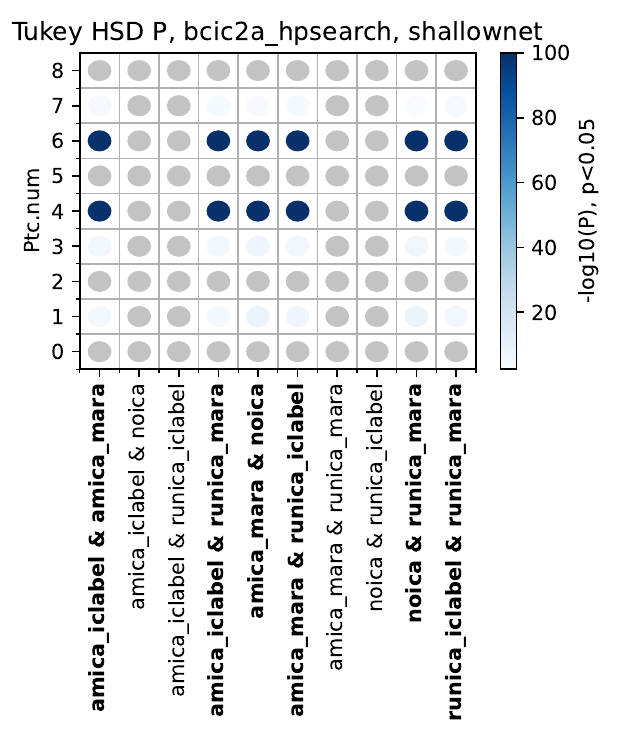}
                b)\includegraphics[width=0.3\textwidth]{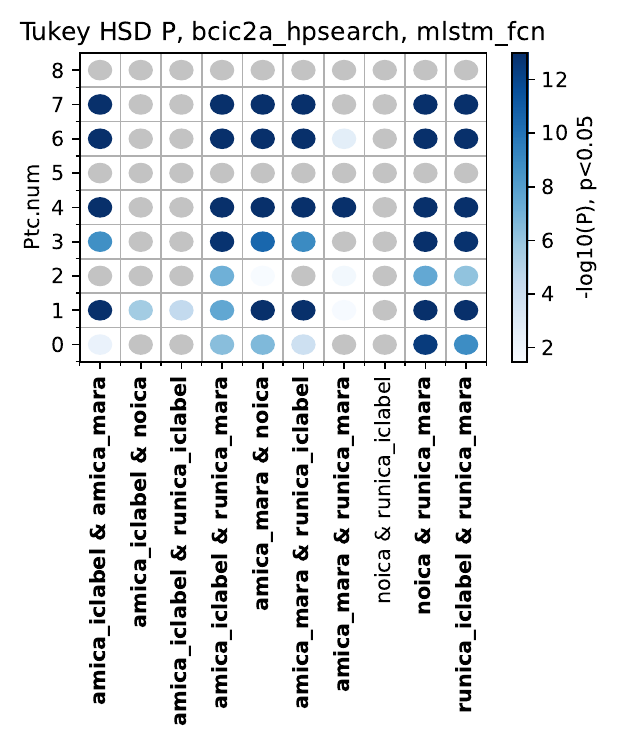}
                c)\includegraphics[width=0.3\textwidth]{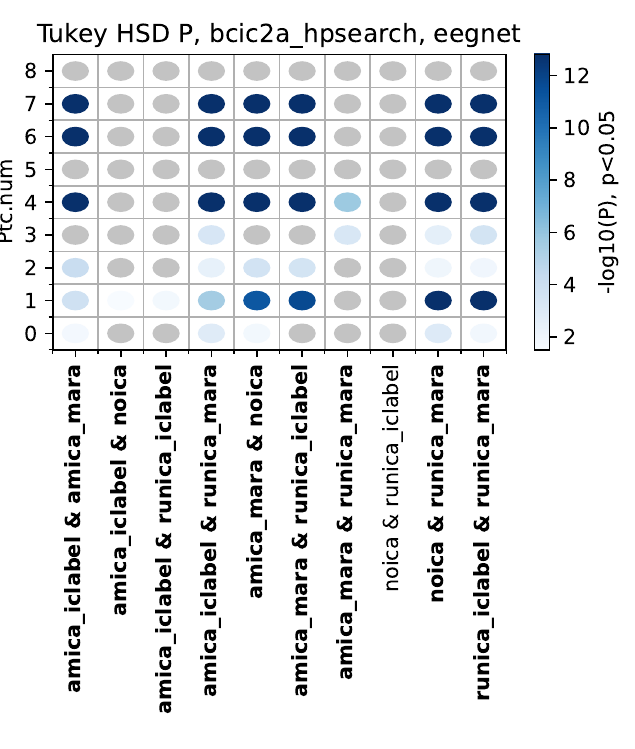}
                	
	\caption{Visualization of post-hoc Tukey's HSD test performed on each participant's entire gridsearch space validation F1 for a) Shallownet, b) MLSTM-FCN, c) EEGNet architectures on the BCIC-IV2a dataset. Each row denotes individual participant in the dataset, while each column represents a single comparison pair among the 5 ICA component rejection pipelines. Cells are shaded in gray if the pair comparison did not yield $p<0.05$. For pairs with significant p-values, cells are shaded by colors representing the negative base-10 logarithmic values of the p.}
	\label{fig:tukeyhsdindiv_bcic42a}
\end{figure*}

\begin{figure*}[htbp]	
	\centering
	\captionsetup{justification=centering}
		a)\includegraphics[width=0.3\textwidth]{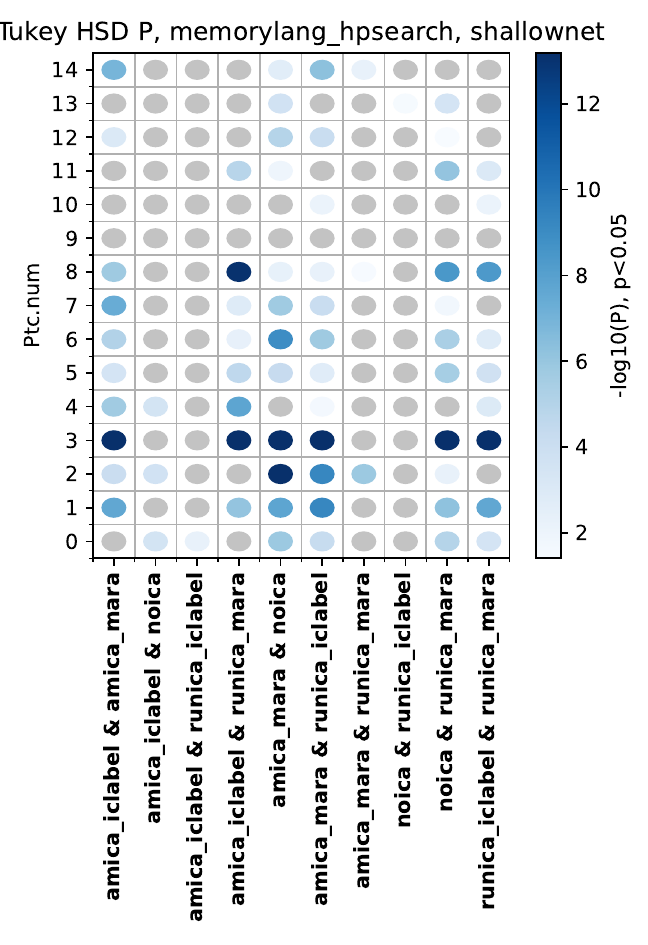}
                b)\includegraphics[width=0.3\textwidth]{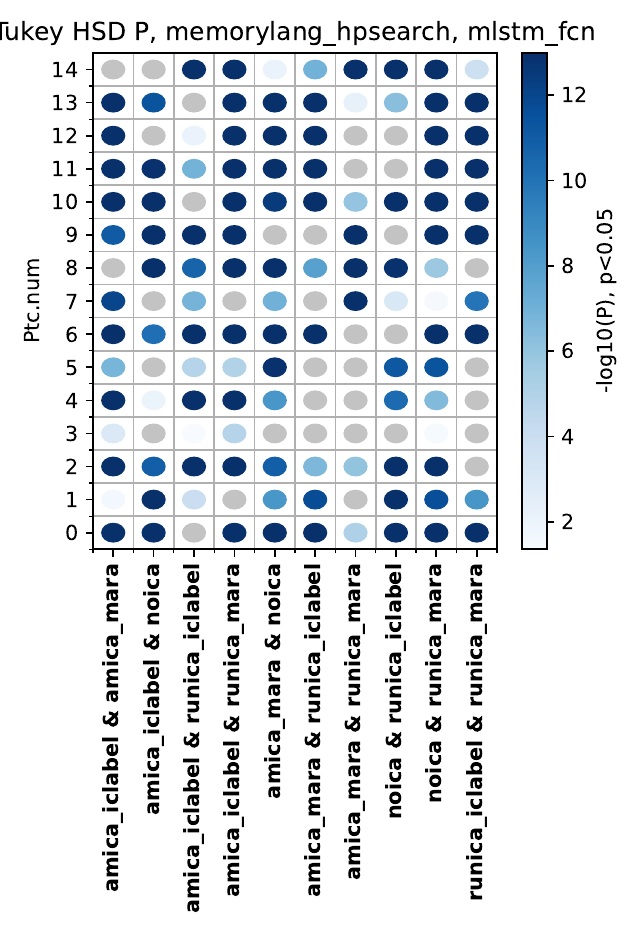}
                c)\includegraphics[width=0.3\textwidth]{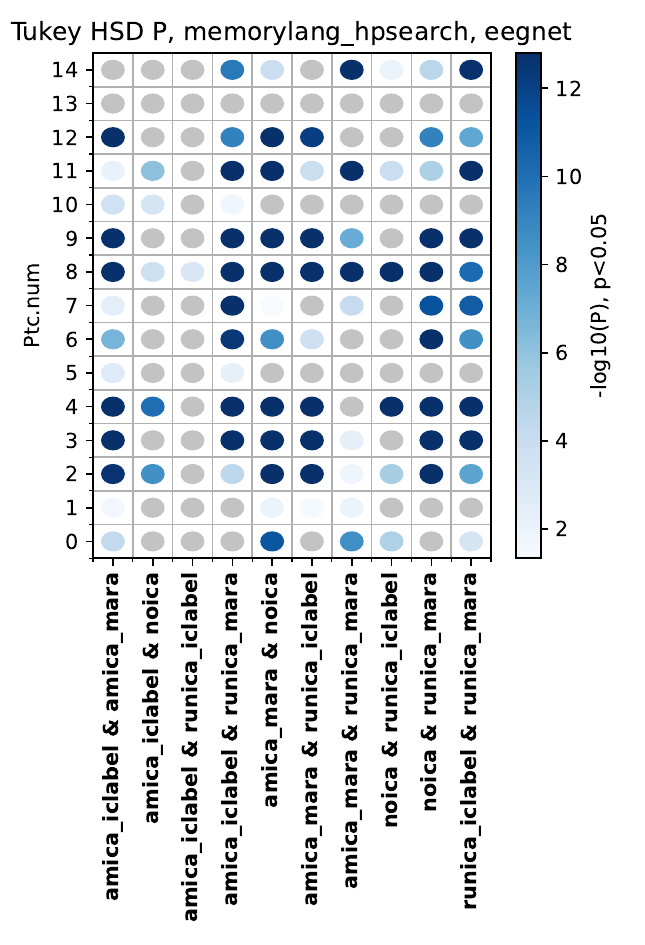}	
	\caption{Visualization of post-hoc Tukey's HSD test performed on each participant's entire gridsearch space validation F1 for a) Shallownet, b) MLSTM-FCN, c) EEGNet architectures on the long-term memory dataset. Each row denotes individual participant in the dataset, while each column represents a single comparison pair among the 5 ICA component rejection pipelines. Cells are shaded in gray if the pair comparison did not yield $p<0.05$. For pairs with significant p-values, cells are shaded by colors representing the negative base-10 logarithmic values of the p.}
	\label{fig:tukeyhsdindiv_memorylang}
\end{figure*}

\begin{figure*}[htbp]
        \centering
        \captionsetup{justification=centering}
                a)\includegraphics[width=0.3\textwidth]{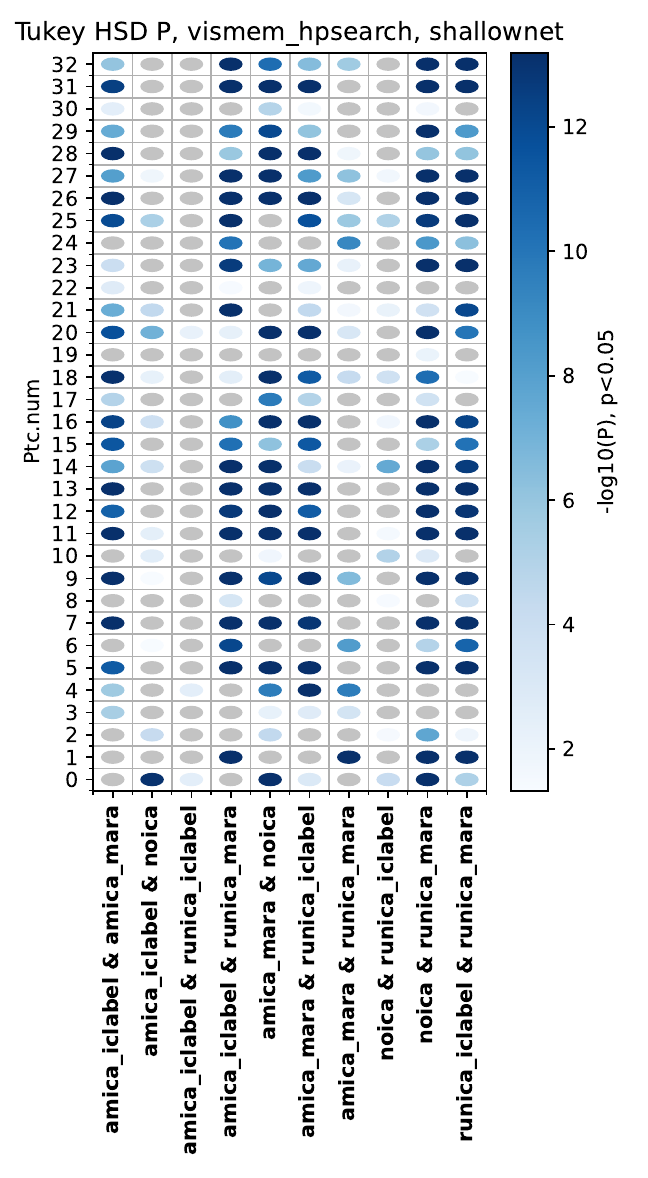}   
                b)\includegraphics[width=0.3\textwidth]{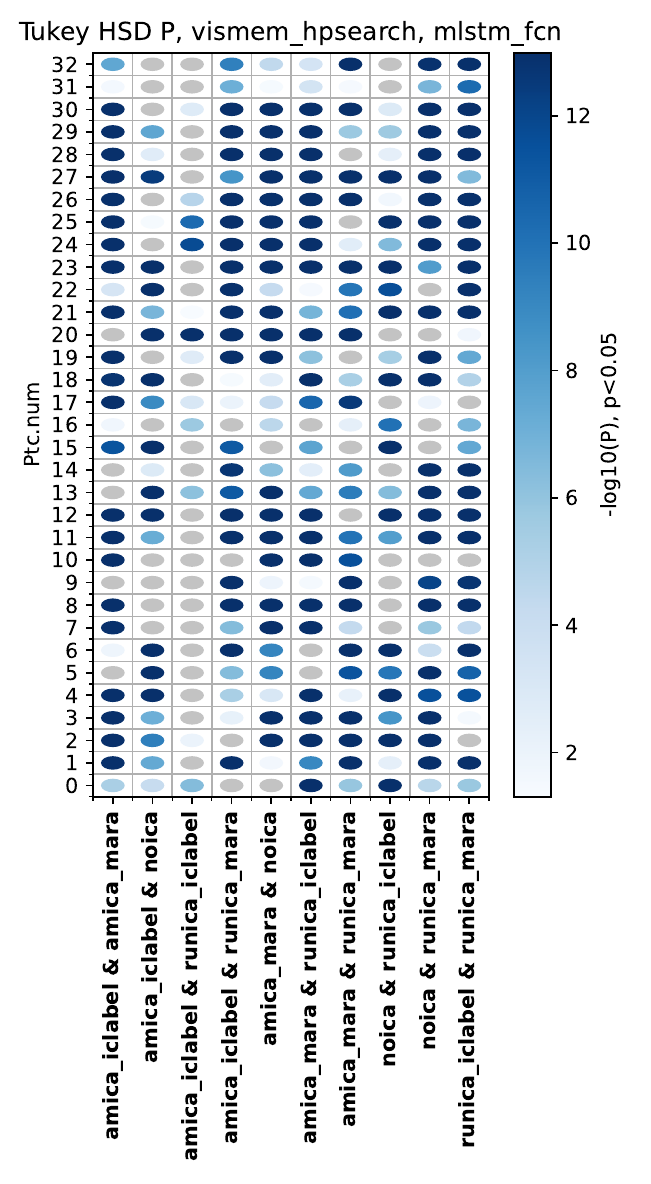}
                c)\includegraphics[width=0.3\textwidth]{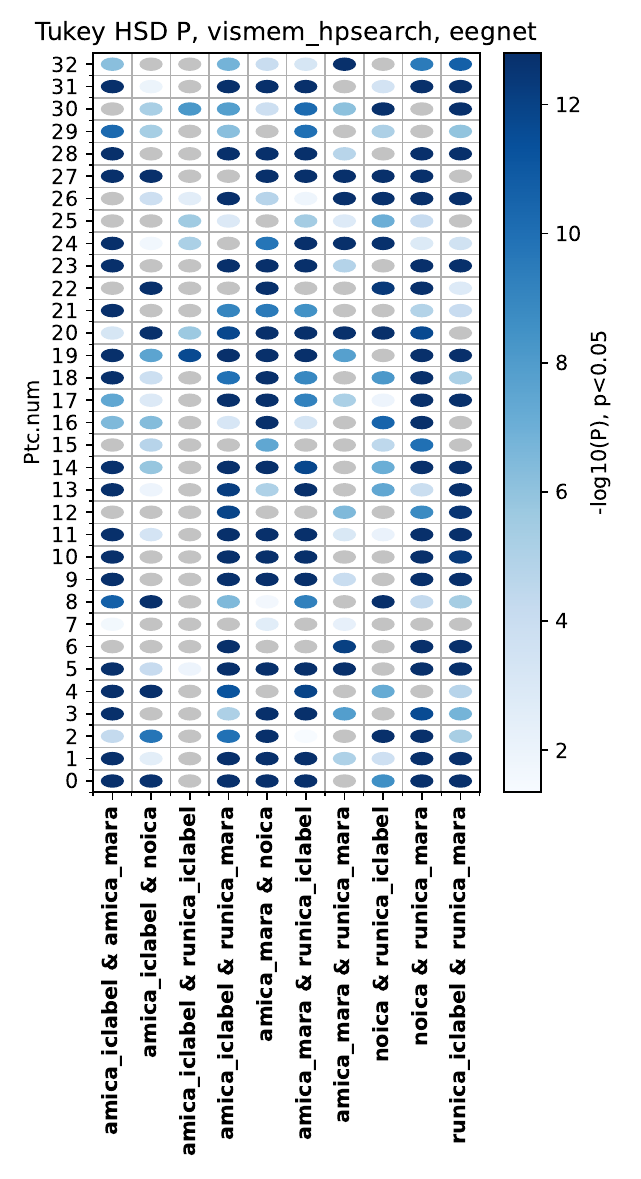}     
        \caption{Visualization of post-hoc Tukey's HSD test performed on each participant's entire gridsearch space validation F1 for a) Shallownet, b) MLSTM-FCN, c) EEGNet architectures on the visual memory dataset. Each row denotes individual participant in the dataset, while each column represents a single comparison pair among the 5 ICA component rejection pipelines. Cells are shaded in gray if the pair comparison did not yield $p<0.05$. For pairs with significant p-values, cells are shaded by colors representing the negative base-10 logarithmic values of the p.}
        \label{fig:tukeyhsdindiv_vismem}
\end{figure*}

Post-hoc Tukey's HSD tests showed significant differences in line with the initial one-way ANOVA results (Figure~\ref{fig:tukeyhsdindiv_bcic42a} for the BCIC-IV2a dataset, Figure~\ref{fig:tukeyhsdindiv_memorylang} for the long-term memory , Figure~\ref{fig:tukeyhsdindiv_vismem} for the visual memory dataset). Notably in the BCIC-IV2a dataset results, for all networks and participants with significant p values, differences were consistently present on all comparison pairs between ICLabel vs. MARA pipelines, and comparisons between NoICA vs MARA pipelines. In the long-term memory dataset, there was no such seemingly apparent pattern of consistently significant pipeline pairs. However, in post-hoc results from ShallowNet, significant differences were present across nearly all the participants with significant ANOVA results in comparisons of (AMICA+MARA vs. RunICA+ICLabel; 12 out of 15 participants) and (AMICA+MARA vs. NoICA; 12 out of 15). In MLSTM-FCN, similar consistently apparent significance was found (RunICA+MARA vs. AMICA+ICLabel; 13 out of 15 participants), (RunICA+MARA vs. NoICA; 15 out of 15), (AMICA+MARA vs. AMICA+ICLabel; 13 out of 15), (AMICA+MARA vs. NoICA; 13 out of 15 participants) as well. In EEGNetV4, highly consistently significant difference was found between (AMICA+ICLabel vs. AMICA+MARA; 13 out of 15 participants), (AMICA+ICLabel vs. RUNICA+MARA; 12 out of 15), and (AMICA+MARA vs. NoICA; 12 out of 15). In all architectures, it appeared the strongest apparently significant differences were between some form of MARA-processed pipelines and the other pipelines. Post-hoc results from the visual-memory dataset showed a similar pattern in which comparison of scores from component rejection methods other than MARA (noICA, ICLabel) with itself yielded more significant difference than otherwise, with their difference being more pronounced in MLSTM-FCN than in other networks.

\begin{figure*}[htbp]
	\centering
		\subfloat[Best Param Validation F1]{
			\includegraphics[width=0.45\textwidth]{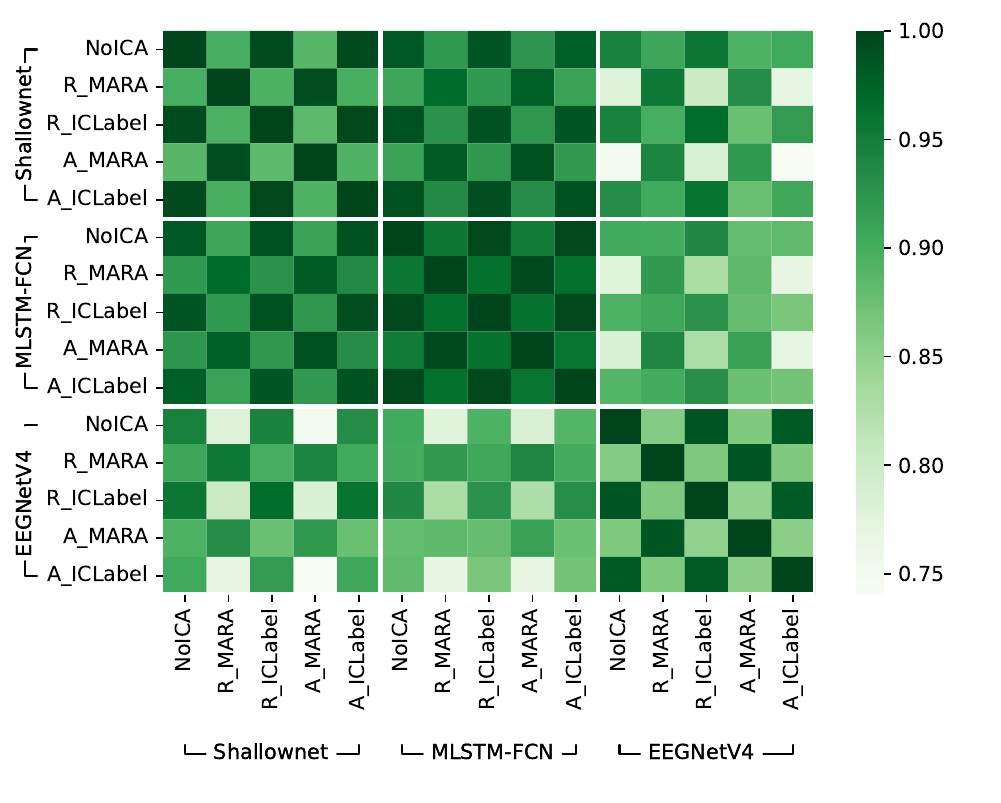}
			\label{subf-f1corr-bestval}
		}
		\subfloat[Test F1]{
			\includegraphics[width=0.45\textwidth]{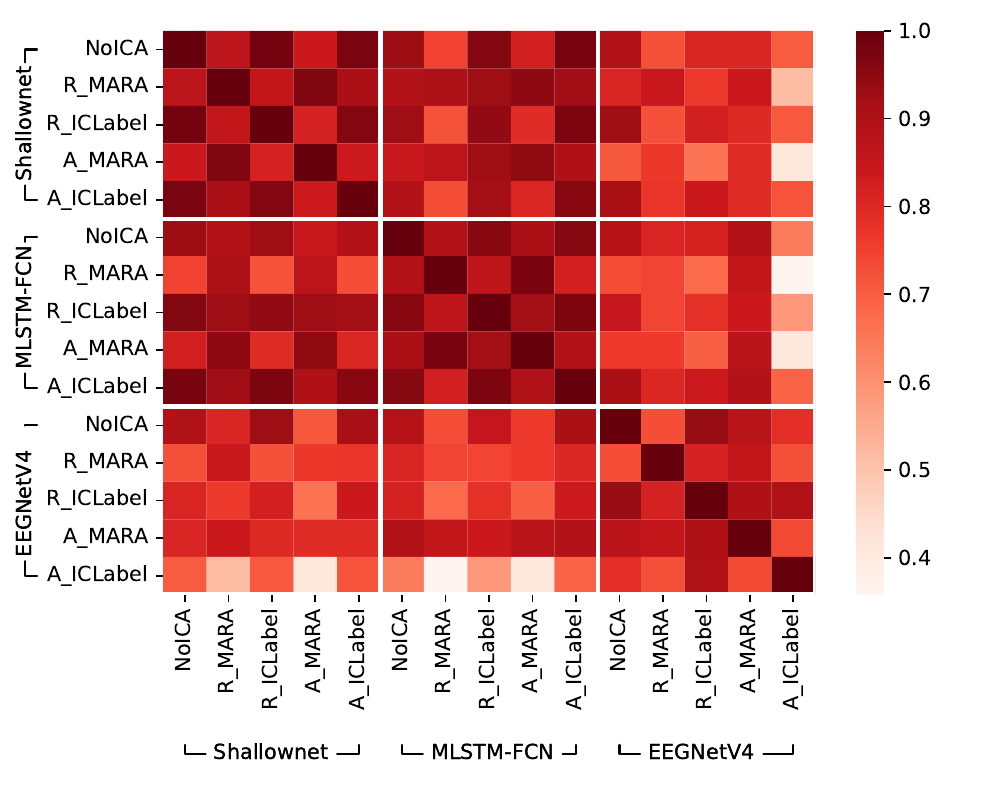}
			\label{subf-f1corr-test}
		}
		
	\caption{Across network and pipelines correlational matrix of the best-parameter validation F1 scores (green) and holdout test-set F1 scores (red) for the BCIC-IV2a dataset. Pipeline labels are abbreviated for space.}
	\label{fig:f1corr_bcic42a}
\end{figure*}

\begin{figure*}[htbp]
	\centering
		\subfloat[Best Param Validation F1]{
			\includegraphics[width=0.45\textwidth]{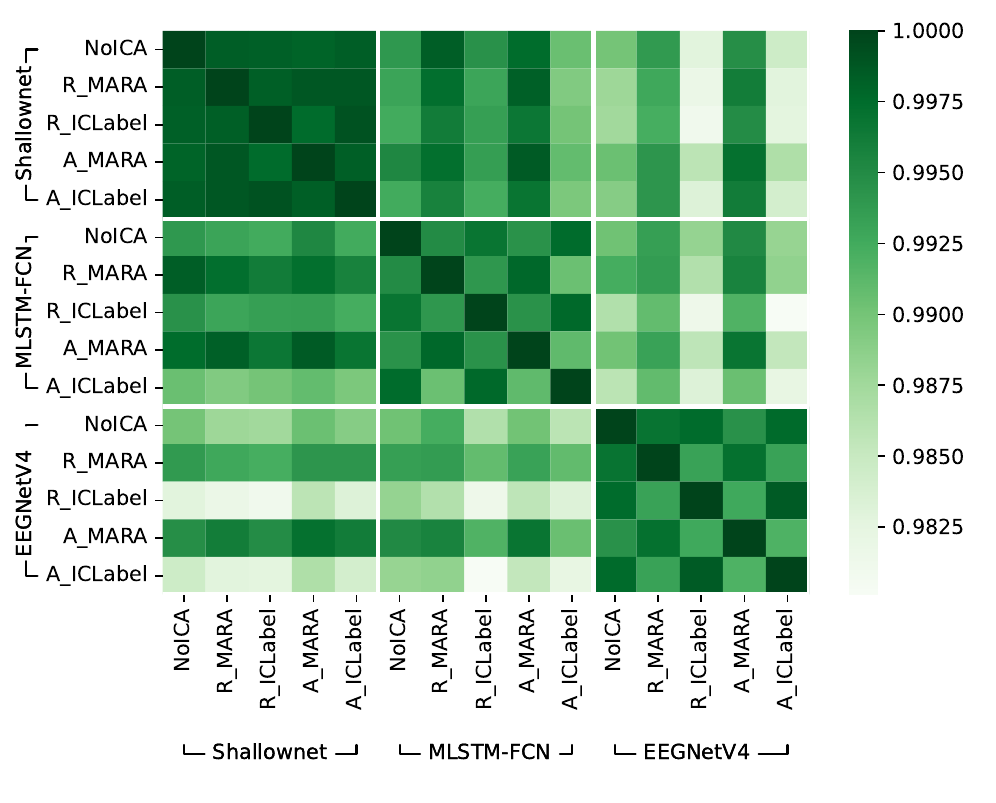}
			\label{subf-f1corr-bestval-memory}
		}
		\subfloat[Test F1]{
			\includegraphics[width=0.45\textwidth]{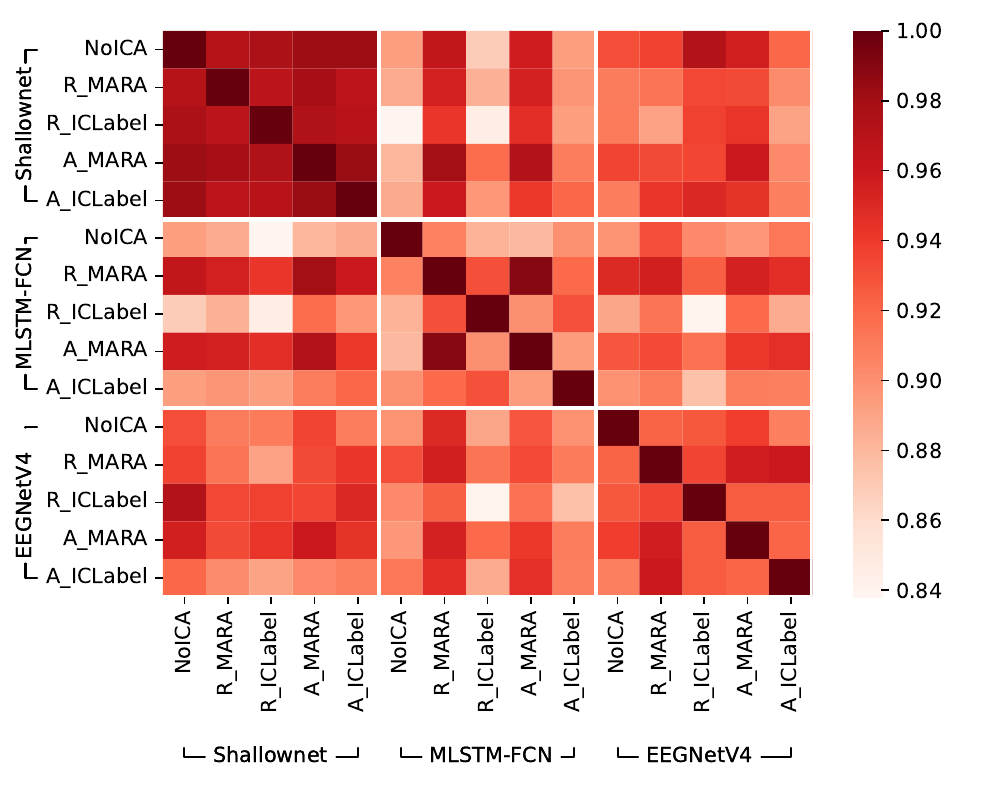}
			\label{subf-f1corr-test-memory}
		}
		
	\caption{Across network and pipelines correlational matrix of the best-parameter validation F1 scores (green) and holdout test-set F1 scores (red) for the long-term memory dataset. Pipeline labels are abbreviated for space. }
	\label{fig:f1corr_memory}
\end{figure*}

\begin{figure*}[htbp]
	\centering
		\subfloat[Best Param Validation F1]{
			\includegraphics[width=0.45\textwidth]{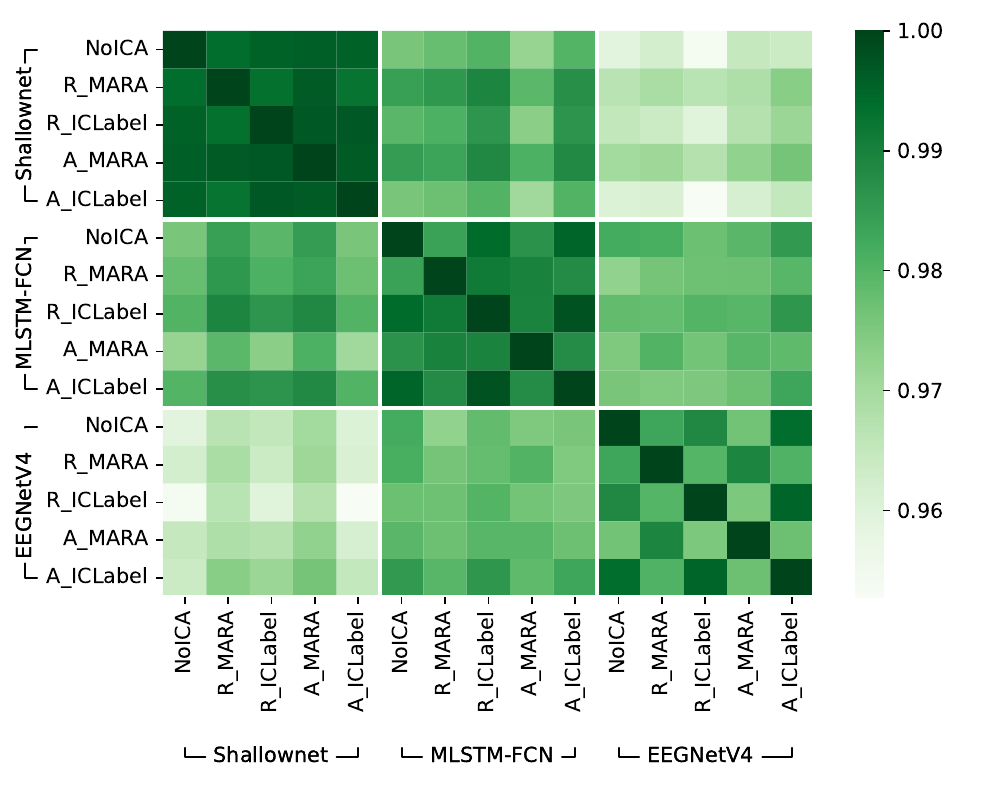}
			\label{subf-f1corr-bestval-vismem}
		}
		\subfloat[Test F1]{
			\includegraphics[width=0.45\textwidth]{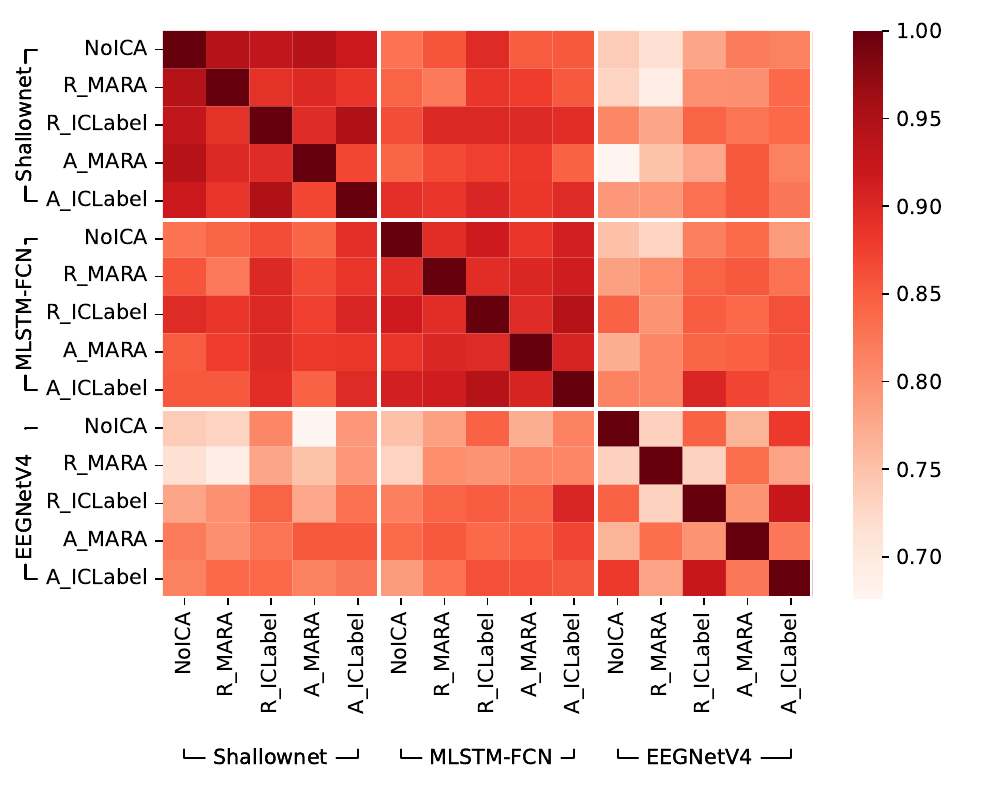}
			\label{subf-f1corr-test-vismem}
		}
		
	\caption{Across network and pipelines correlational matrix of the best-parameter validation F1 scores (green) and holdout test-set F1 scores (red) for the visual memory dataset. Pipeline labels are abbreviated for space. }
	\label{fig:f1corr_vismem}
\end{figure*}

Supplementary Material Table 7 and 8 show correlation matrices between the pipelines for each network and dataset of all validation F1 scores from the entire grid search.
Table~\ref{tab:val_test_corr} shows the calculated correlation between Validation F1 scores from the best performing hyperparameter set and the final hold-out test-set F1 scores (df=6 for BCIC-IV2a dataset, df=12 for long-term memory dataset). Evidently the validation scores and the final test-set scores were highly correlated.  Figure~\ref{fig:f1corr_bcic42a},  ~\ref{fig:f1corr_memory}, and ~\ref{fig:f1corr_vismem} show correlation matrices for each dataset across ICA pipelines as well as across different networks. Classifier performance from individual participant data was again highly correlated for different pipelines and even different networks. In the BCIC-IV2a dataset, alternating patterns in which the MARA-based rejection pipelines had comparably lower correlation to other pipelines can be see in both validation F1 and Test F1 correlation matrices. Correlations within F1 scores of EEGNetV4 were higher than the network's correlations with other architectures. In the long-term memory dataset, the difference of correlations with EEGNet and other architectures was especially pronounced in the pipelines rejected with ICLabel for the validation F1, although the pattern was less apparent in the Test F1 correlations. In both validation and test F1, pipeline results within Shallownet architecture had the highest correlations compared to across and within other networks for both validation and test F1. In the visual memory dataset as well, correlations between performances of different networks showed EEGNet architecture being the least correlated, although the baseline was much higher than in results from the BCIC-IV2a dataset.

Supplementary Figures 16 (BCIC-IV2a dataset), 17 (long-term memory dataset), and 18 (visual-memory dataset) show best-parameter validation F1 and final test-set F1 scores of different pipelines side by side, grouped for each participant. Based on the one-way ANOVA and the post-hoc Tukey tests performed on individual participant level, while the majority of participants had significantly different validation set performance across different hyperparameter setups, it can be seen that the best validation performance, along with the final holdout test-set performance do not seem to vary much between different pipelines in the same network. 
This data shows that performance from the same participant in reasonably optimized networks is largely consistent across pipelines, and in fact origin (the participant) of the data seems to have a much stronger influence on performance compared to the pre-processing pipelines themselves.

\begin{table*}
	\scriptsize	
	\captionsetup{justification=centering}
	\centering
	\begin{center}
			\begin{tabular}{|l|c|c|c|}				
				\hline	
				\multicolumn{4}{|c|}{Best Validation F1 \& Test F1 correlation R(P val, CI)}\\
				\hline
				Dataset|Networks&ShallowNet&MLSTM-FCN&EEGNetV4\\
				\hline
				BCIC-IV2A&$.931(p<1e^{-19}, CI:[.878, .962])$&$.897(p<1e^{-16}, CI:[.819, .942])$&$.683(p<1e^{-6}, CI:[.487, .813])$\\
				\hline
				Long-Term Memory&$.976(p<1e^{-49}, CI:[.963, .985])$&$.923(p<1e^{-31}, CI:[.880, .951])$&$.935(p<1e^{-33}, CI:[.899, .959])$\\
				\hline
				Visual Memory&$.943(p<1e^{-80}, CI:[.924, .958])$&$.913(p<1e^{-65}, CI:[.884, .935])$&$.853(p<1e^{-48}, CI:[.806, .890])$\\
				
				%
				%
				%
				\hline			
			\end{tabular}
	\end{center}

	\caption{Correlation between best performing validation set's F1 scores and the final holdout test set F1 scores, for each dataset and network.} 
	\label{tab:val_test_corr}	
\end{table*}

\subsection{Comparison of classification accuracy across pipelines: Group level }
\begin{figure*}[htbp]
	\centering
	\captionsetup{justification=centering}
		a)\includegraphics[width=0.3\textwidth]{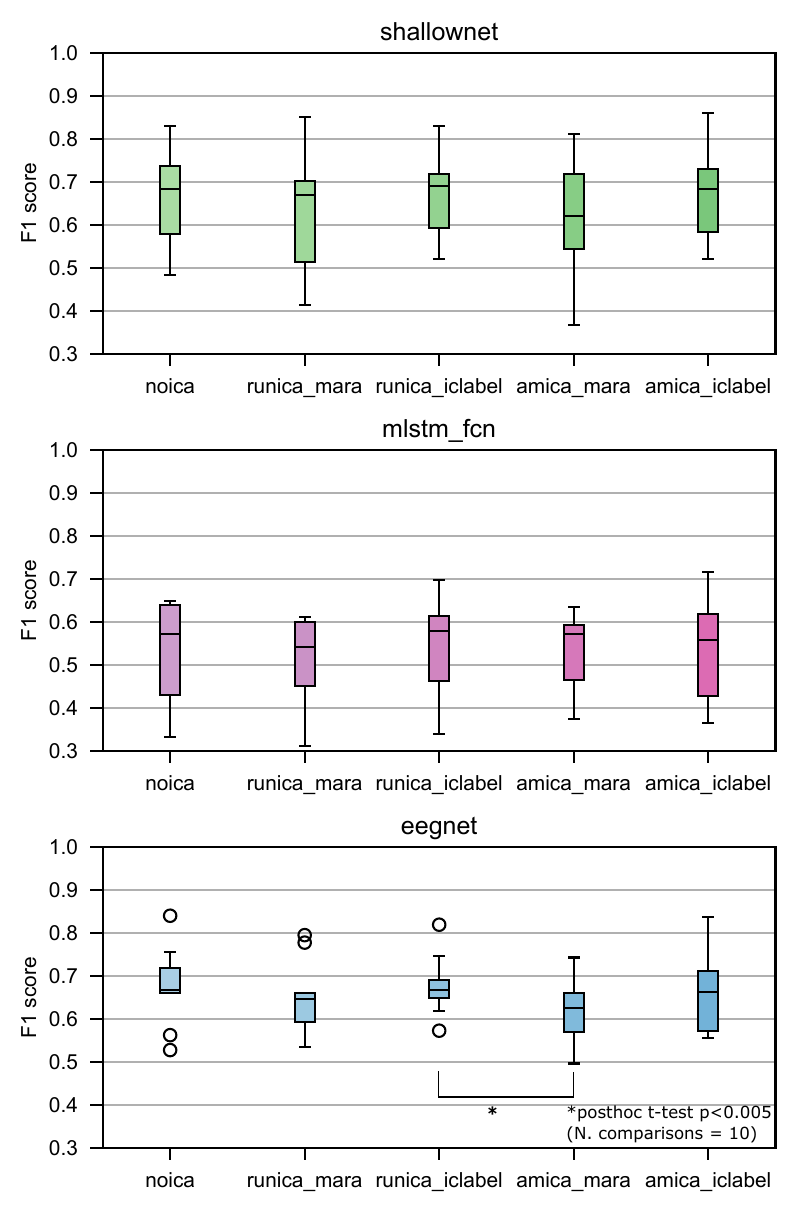}
                b)\includegraphics[width=0.3\textwidth]{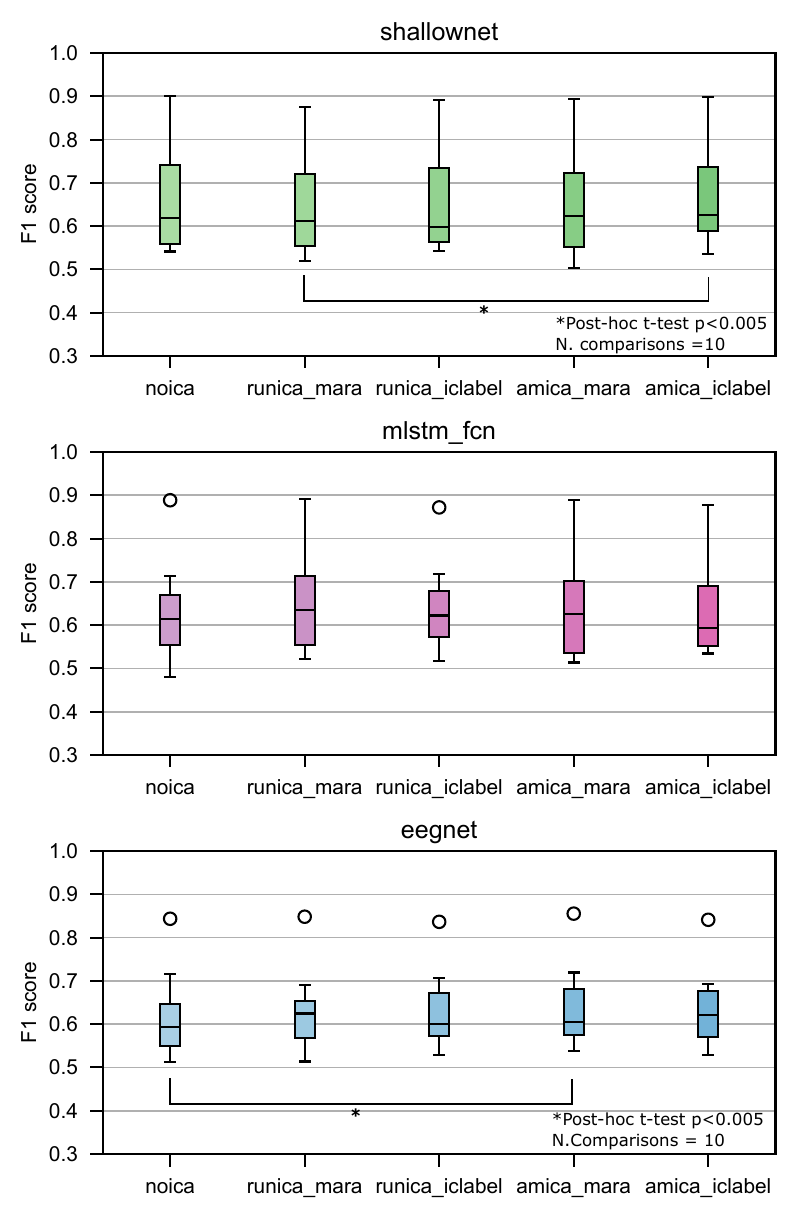}
                c)\includegraphics[width=0.3\textwidth]{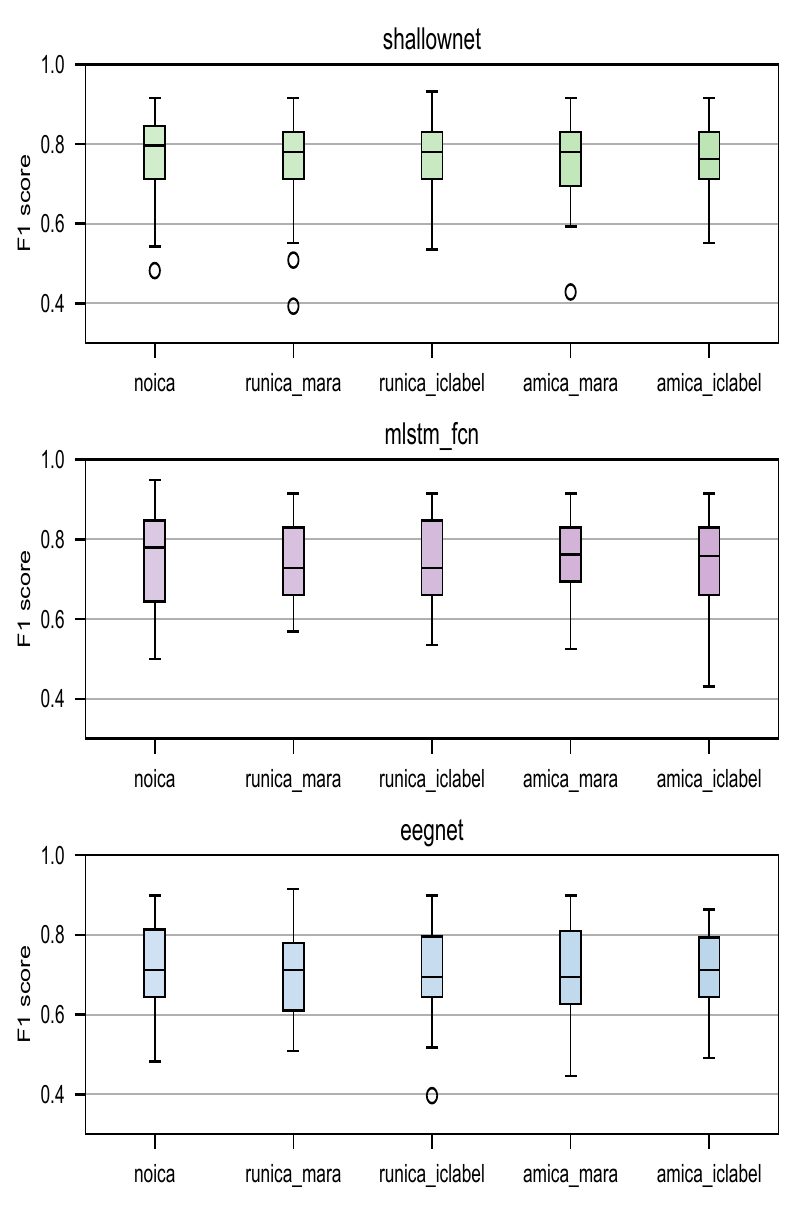}
	\caption{Overall test-set F1 score boxplots for each pipeline for each network for a) BCIC-IV2a, b) Long-term memory, c) Visual memory datasets. Significance markings of pipeline comparisons are based on post-hoc paired T-tests corrected with Bonferroni correction where the corrected p-value was lower than 0.05. Whiskers are defined as higher and lower quantiles of the data, and the center line denotes the median. Outlier values are denoted with circles.}
	\label{fig:lvl2_testf1_box}
\end{figure*}

Figure~\ref{fig:lvl2_testf1_box} shows the comparison of pipeline performances on a whole-dataset level across participants. In this analysis, only holdout \emph{test-set F1 scores} were considered for a repeated-measure analysis of variance (RM-ANOVA) where each participant was counted as a repeated measure. Compared to one-way ANOVAs performed on the participant individual level, observed differences between the pipelines were much weaker. In the initial RM-ANOVA, p-values were lower than 0.05 for ShallowNet and EEGNetV4 while not in MLSTM-FCN, for both BCIC-IV2a (ShallowNet F($df_{grp} 4, df_{rm} 32$)=2.728, p=.046; MLSTM-FCN F=1.150, p=.351; EEGNetV4 F=3.096, p=.029) and long-term memory dataset (ShallowNet F($df_{grp} 4, df_{rm} 56$)=5.432, p=.001; MLSTM-FCN F=1.151, p=.338, EEGNetV4 F=3.203, p=.019). In a post-hoc analysis with paired T-tests corrected by Bonferroni correction (N comparisons = 10), the difference was found to be only significant between RunICA+ICLabel and AMICA+MARA pipelines in results trained from EEGNetV4 for the BCIC-IV2a dataset, while in the long-term memory dataset significant differences were found between RunICA+MARA and AMICA+ICLabel for ShallowNet, and between NoICA and AMICA+MARA in EEGNetV4. In results from the visual-memory dataset, no significant differences were observed from the RM ANOVA at all. See Supplementary Tables 10 for BCIC-IV2a dataset, 11 for the long-term memory dataset, and 12 for the visual memory dataset in specific post-hoc T-test results.

\subsection{Number of rejected components and classifier performance}
\begin{figure*}[htbp]
	\centering
	\captionsetup{justification=centering}
		\includegraphics[width=1\textwidth]{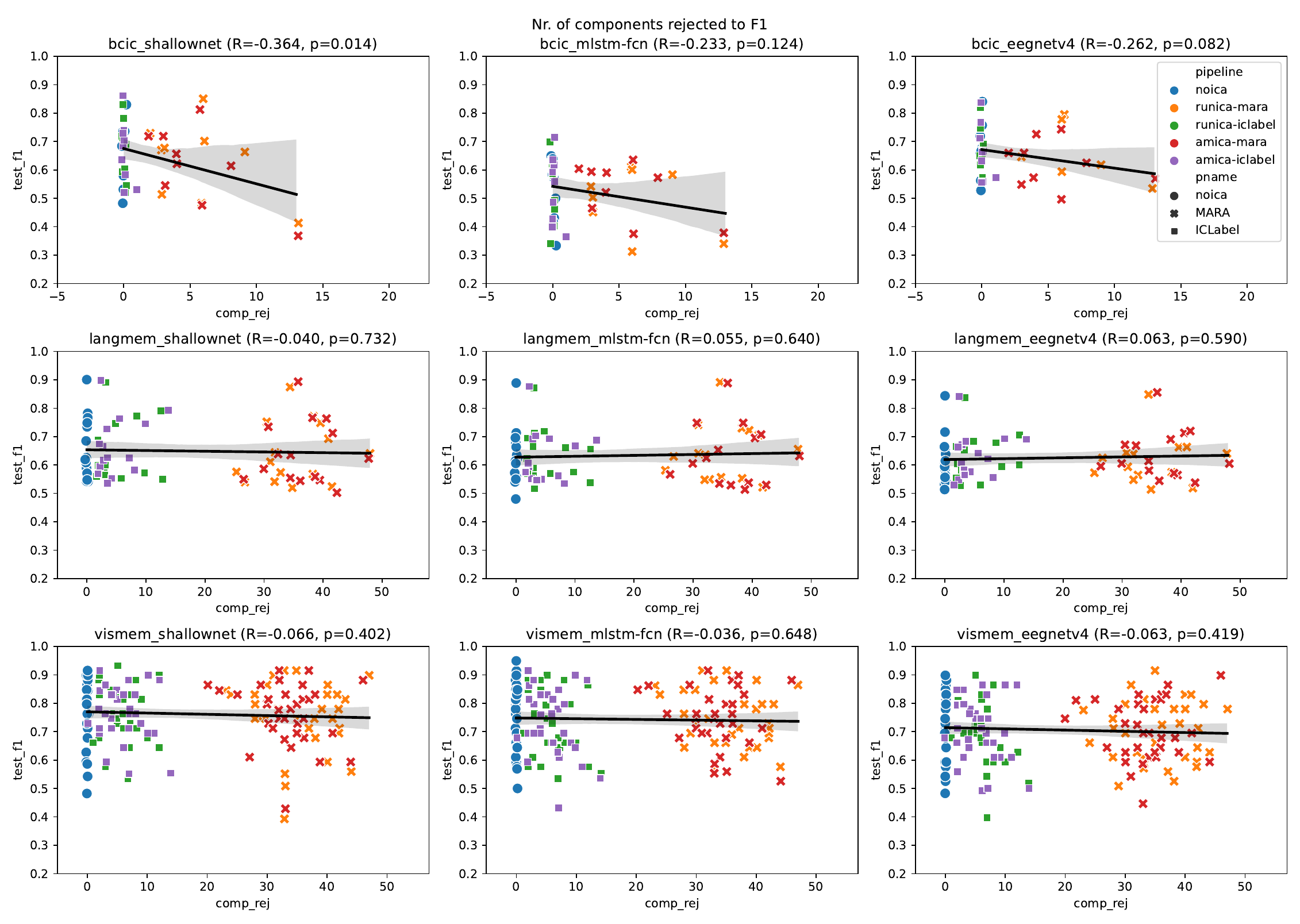}
	\caption{Scatterplots of number of components rejected and the test-F1 scores in each dataset-model combination given for each participant in dataset. X-axis shows the number of components rejected as a result of ICA-based noise rejections, Y-axis shows the test-F1 scores of each participant. The X-values are slightly jittered for visual readability as noICA pipelines all have 0 for component rejection values. Row orders denote the dataset: the BCIC-IV2a dataset, lanugage-memory dataset, and the visual memory dataset. Column orders denote model architecture: ShallowNet, MLSTM-FCN, and EEGNetV4. The denoted p-values are not multiple comparison corrected.}
	\label{fig:comp_to_f1_plot}
\end{figure*}

To see whether the number of rejected components correlate to changes in classifier performance, we next performed a correlation analysis between the rejected component counts from each pipeline-participant to the test-set F1 scores from the best-performing models of each architecture. Figure~\ref{fig:comp_to_f1_plot} shows a scatterplot of individual participant test-F1 values from different pipelines for each network and dataset setup, plotted against the number of components rejected. In each dataset-architecture combination, Pearson's R values were also calculated, shown on the plot subtitles. The displayed p-values are not multiple-comparison corrected, and none of the p-values survived Bonferroni correction. In the BCIC-IV2a dataset, a close-to-significant negative correlation was found between the number of rejected components and test-set F1 scores ($R(Shallow)=-.364, p_{uncorr}=.014; R(MLSTM)=-.233, p_{uncorr}=.124; R(EEGNet)=-.262, p_{uncorr}=.082$).

Building up from the above consideration of the number of components rejected in all datasets, we see a further interesting pattern: In the BCIC-IV2a dataset, there was a difference of 5 rejected components on average between the MARA and AMICA pipeline out of 22 independent components. In the long-term memory dataset, the difference was even more stark with up to 30 components difference between the two rejection pipelines out of 62 independent components. Comparable differences in rejected components between MARA and ICLabel were found in the visual memory dataset as well. Importantly, however, despite the large gap in remaining components between the component rejection methods, we found no consistent, similar difference in classifier performance. Combined with the observation that performance metrics between different pipelines were highly correlated, it seems possible that while noise does exist in the different datasets (especially in the long-term memory dataset, which had a much longer recording time), this does not seem to affect the deep neural networks performance in highlighting the features relevant to the task, nor are noise features recognized as relevant to the classification at all.

Even in participants with larger performance differences between pipelines (e.g. participant 5 in BCIC-IV2a dataset, participant 15 in long-term memory dataset, participant 2 in visual memory dataset), we were unable to pinpoint a consistently best-performing pipeline. Additionally, the NoICA pipeline (in which no component rejection was performed at all) is \emph{not} the worst performing pipeline in these comparisons. Furthermore, for several participants displaying clear performance differences (such as participant 2, 5, 7 in the BCIC-IV2a dataset), pipelines processed with MARA resulted in worst overall performance. Although this may be due to the significantly larger number of components rejected in the MARA pipeline overall, the fact that the performance dip is more apparent in the BCIC-IV2a dataset (with a generally smaller number of components rejected regardless of the pipeline) compared to the long-term memory dataset is noteworthy.

\section{Discussion}
In this study, we pre-processed 3 datasets of epoched EEG signals purposed for BCI classification through a total of 4 (+1 control) different ICA-based automatic pipelines for rejecting artifact components, to investigate whether such pre-processing procedures affect general classifier performance in neural network-based architectures. To see whether the results would be generalizable, we chose 3 different network architectures, including two specifically designed for brain signal classification. Furthermore we optimized hyperparameters in two stages: first parameters that could be shared across different architectures (such as learning rate) were optimized for all 3 networks on one benchmark dataset (Motor Imagery dataset from the BCI competition). Then we optimized network-specific hyperparameters individually for each participant in each dataset, for each network, for each ICA component rejection pipeline, each with a predefined network and dataset-specific parameter search space (that was identical across participants and pipelines) - this resulted in over 800 separate runs of the hyperparameter grid search, each run with a search space of at least a hundred possible combination of parameters. While one-way ANOVA results did reveal some significant differences in classifier performance between ICA pipelines for each individual participant, and more weakly in an across-participant level RM-ANOVA as well, questions remain on whether the ICA based component rejection process is truly beneficial in BCI classifier performance improvement.

\subsection{Is ICA component rejection really beneficial for classifier performance boost?}
Despite the observed significant performance difference between different input pipeline data apparent in the majority of participants across networks and datasets on an individual participant level, the overall benefit from different ICA component rejection pipelines is not as clear when it comes to final performance defining metrics. While the difference between ICA pipelines was pronounced for less-optimized hyperparameter setups of individual participants, once a reasonably well performing parameter was found, the resultant test performance was largely dissimilar. This is observable in Figure~\ref{fig:lvl2_testf1_box}, as well as in Supplementary Figures 16, 17, 18.

\begin{figure*}[htbp]
	\centering
	\captionsetup{justification=centering}
		\includegraphics[width=0.95\textwidth]{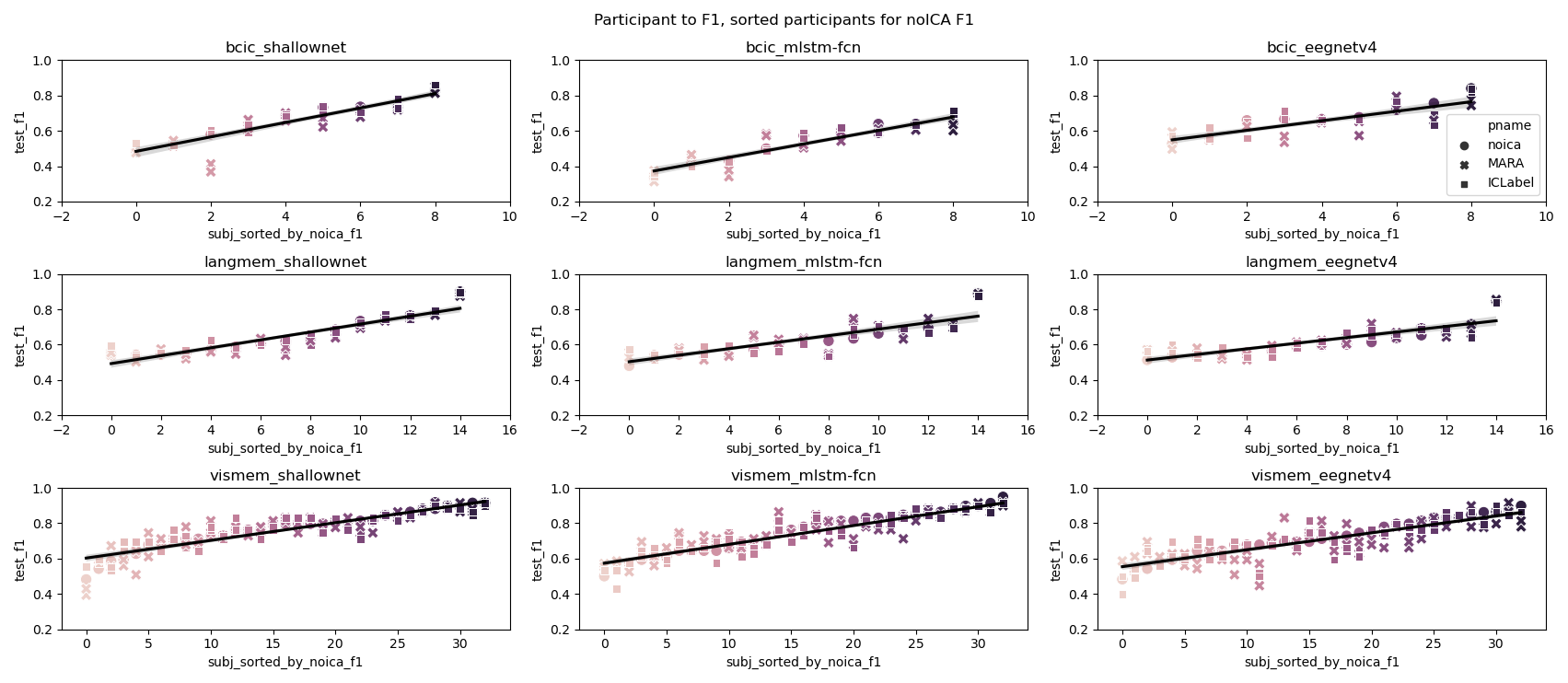}
	\caption{Participant F1 scores for all pipelines, with participant number re-sorted by NoICA pipeline F1 scores.}
	\label{fig:noica_sort_participant_f1}
\end{figure*}

Considering the results so far, we suggest that there is little to no benefit to be gained from utilizing ICA-based noise removal in EEG data: this is true both at the individual participant-dependent level classification, and especially more so on across-participant level. Based on the results of comparing rejected number of components with actual test-set F1 performance, it appears that removing components even for noise rejection purposes has little benefit for network performance and may sometimes be detrimental. ICA computations are generally regarded as computationally expensive in comparison to other methods in the pre-processing pipeline. Especially in multi-dataset prediction studies where the focus is on training inference models rather than performing univariate statistical analysis of EEG epochs, it may be worth reconsidering whether it's truly necessary to apply such methods for a rather inconsistent and possibly marginal boost to prediction performance.
It should also be noted that at the participant-dependent level, the degree of freedom in one-way ANOVA was in the hundreds, likely amplifying what statistical power there was despite the small effect size. This was much less an issue in dataset-representative level RM-ANOVA, as the degree of freedom was limited to the number of participants available in the dataset. In fact, overall performance appears to be more influenced by the data source (the participant) rather than whichever pipelines. When participant order is re-arranged for test-F1 values of a one pipeline (see Figure~\ref{fig:noica_sort_participant_f1} for participants sorted by NoICA F1 scores), the rest of the pipelines seem to more or less follow a similar pattern as that pipeline. This is also evidenced by the high F1 correlations between pipelines and architectures, as seen in Figures~\ref{fig:f1corr_bcic42a}, \ref{fig:f1corr_memory}, \ref{fig:f1corr_vismem}.

\begin{figure*}[htbp]
	\centering
	\captionsetup{justification=centering}
		\includegraphics[width=0.8\textwidth]{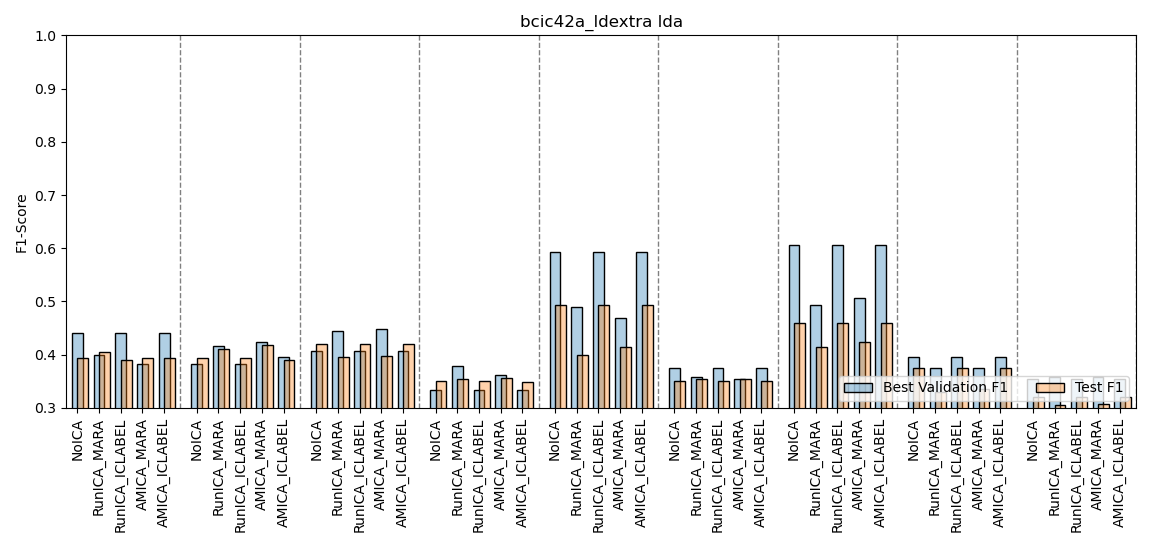}
	\caption{LDA results of the BCICIV2a dataset, performed on raw EEG waveforms without additional feature extraction that are standard for motor imagery.}
	\label{fig:lda_f1_bcic}
\end{figure*}

Are these findings transferable to standard machine learning models? Figure~\ref{fig:lda_f1_bcic} shows individual participant validation and test-f1 scores of LDA models trained on the BCIC-IV2a dataset raw waveforms. Here as well, there is no pipeline that consistently outperforms the others. However, we deem it insufficient to reach similar conclusions as in network models, largely because LDA severely unperforms, with some results close to chance levels. The neural network models that were used in our study, namely the ShallowNet, are reported to have designed intermediate layers after certain feature extraction methods such as filter bank common spatial patterns (FBCSP) ~\cite{schirrmeister2017deep}.
They take raw time-series format EEG as input, and no separate feature extraction methods are necessary for reasonable performance. In standard ML models however, feature extraction is more often performed beforehand rather than using raw waveforms as input, especially in MI datasets. We suspect this may explain the much lower LDA performance with raw data, but also realize adding the feature extraction step for the standard ML models will complicate establishing equal comparison with the network models, as the data technically undergoes an additional layer of processing through feature extraction. As such, we believe extending the findings to standard ML is out of the current scope of the study.

It is worth noting that so far our study has investigated three different datasets of brain data. In a related study of evaluating the effect of various data cleaning methods on general classification tasks with multiple datasets (not specifically limited to EEG or time series), Li et al. notes that not only is the impact of data cleaning on classifiers hardly consistent across models, but also the impact strongly depends on the dataset: as different datasets have different error distributions, no one cleaning method seems to be consistently superior~\cite{li2021cleanml}. From our current results, however, while clear differences in the number of artifact components can be observed in the number of rejected components across the three datasets, the improvement of performance from ICA rejection methods were negligible in terms of effect size. 
While standard ERP-based analysis results are outside the purview of our study, the lack of benefit from employing ICA based rejection methods appear to be also apparent in such studies. Recent results from Delorme suggest that automated rejection methods for removing muscle and eye movement components do not lead to more channels with statistical significance in ERP~\cite{delorme2023eeg}. 
Regardless, We believe extending our study to more datasets of different BCI tasks will help to get an even clearer picture of whether ICA noise rejection is actually beneficial for classification purposes.

\subsection{Network activation maps across pipelines and EEG waveforms}
\begin{figure*}[htbp]
	\centering
	\captionsetup{justification=centering}
		\includegraphics[width=0.8\textwidth]{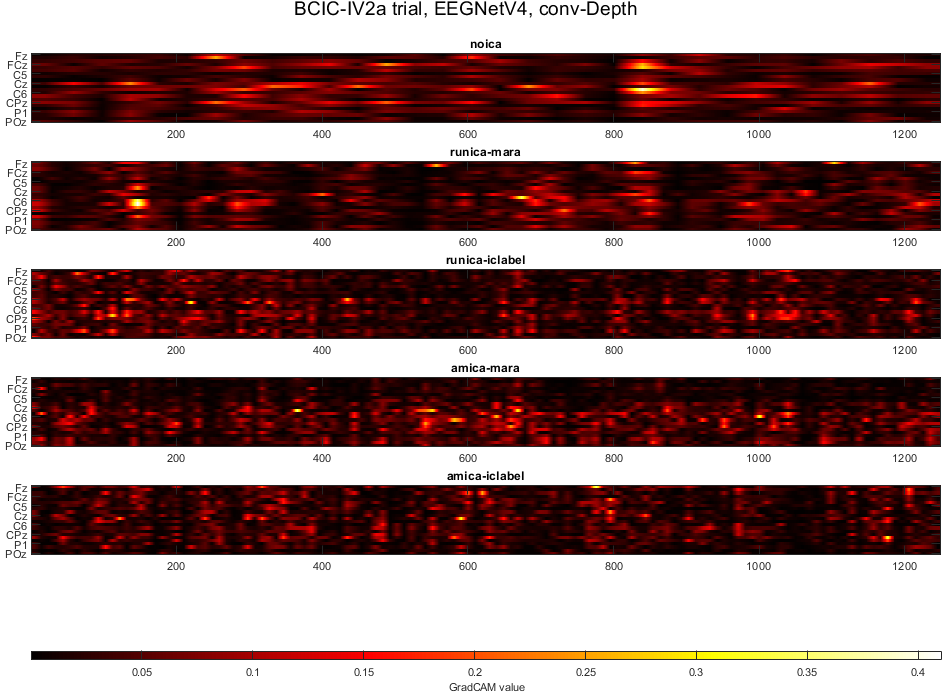}
	\caption{Averaged GradCAM values from the BCICIV2a dataset, participant 1, test set from the EEGNetV4's best parameter model of each pipeline. Values from the conv-separable-depth layer were computed.}
	\label{fig:gradcam_avg}
\end{figure*}

While performance across different pipelines was similar for the same dataset participant, the question remains whether the instances of the same network model from the different pipelines were utilizing similar features for a given sample. Furthermore, there is merit in observing whether output EEG waveforms from the different ICA pipelines were similar in the first place, and how they match up to the class activations of networks. As a preliminary analysis of this question, we extracted Gradient-weighted Class Activation Maps (GradCAM)~\cite{selvaraju2017grad} values from the "separable-depth" convolutional layer of the best performing EEGNetV4\footnote{Usually, activations from the final convolutional layers are used for CAM visualization, and we chose EEGNetV4 for the preliminary analysis, as the top convolutional layer of the Shallownet already has different spatial channel information combined into a single dimension, and MLSTM-FCN's 1D-CNN layer activations seemed to highlight most time dimensions of a given spatial channel as a whole.} models from each pipeline using the test-set data of the first participant in the BCIC-IV2a dataset. The averaged CAM heatmap of the test-set for each pipeline can be found in Figure~\ref{fig:gradcam_avg}.

\begin{figure*}[htbp]
	\centering
	\captionsetup{justification=centering}
		a)\includegraphics[width=0.46\textwidth]{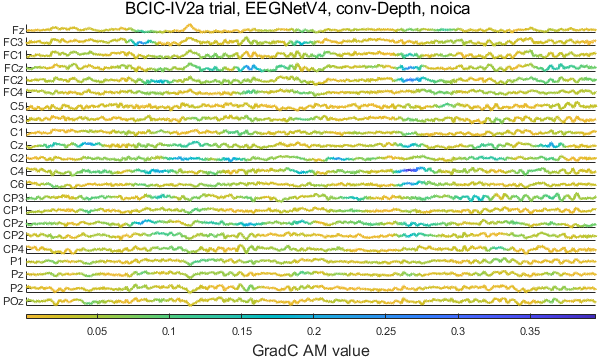}
        b)\includegraphics[width=0.46\textwidth]{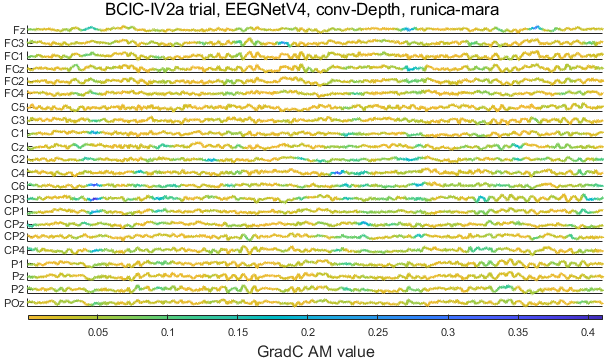}
        c)\includegraphics[width=0.46\textwidth]{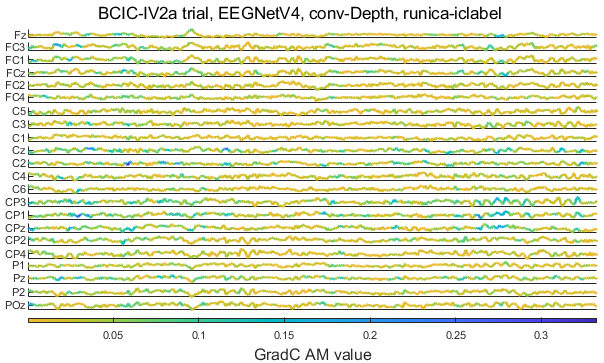}
        d)\includegraphics[width=0.46\textwidth]{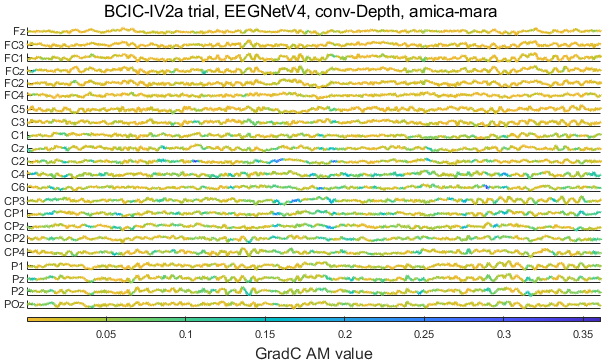}
        e)\includegraphics[width=0.46\textwidth]{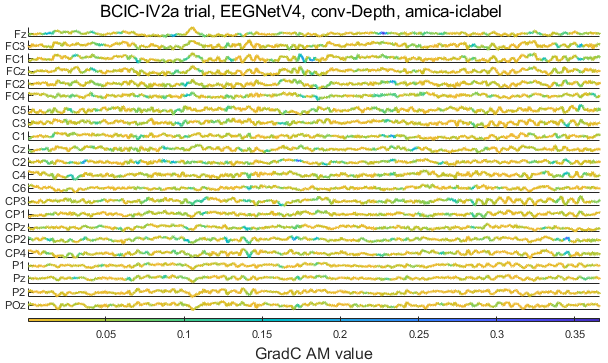}
	\caption{Comparison of Raw EEG waveforms of a test set trial from the BCICIV2a dataset, participant 1, for each pipeline. GradCAM values of the best performing EEGNetV4 model are shown as color gradients on the waveforms.}
	\label{fig:gradcam_trial}
\end{figure*}

GradCAM is a visualization method to assess highlighted features in individual layers, and as such it is not immediately intuitive as to what features are deemed important for time series.
Next, we tried to see how the class activation maps were reflected on EEG trial waveforms. For longer duration epochs, it is rather uncommon to average raw waveforms across trials, and as such we chose a sample trial from a given participant. We superimposed the CAM values to the waveforms in each pipeline, and the result can be seen in Figure~\ref{fig:gradcam_trial}. 
While we are not aware of measures to quantifiably compare different class activation maps, as CAM results (and to an extent EEG data) can be interpreted as a single-channel 2 dimensional image, we computed structural similarity index values (SSIM)~\cite{wang2004image} for each pipeline to determine how similar the activations, as well as the output waveforms from individual ICA pipelines were to one another. Table~\ref{tab:cam_ssim} shows the SSIM values from EEG epochs of the entire test-set of participant 1 from the BCIC dataset in each pipeline, as well as that of the CAM values from the same data in EEGNetV4. In the EEG waveforms, we see that NoICA and the two ICLabel pipelines produce very similar values to each other, and the two MARA pipelines are also similar to each other. This is expected given the number of rejected components were the same for ICLabel and NoICA in the BCIC dataset, and the two MARA pipelines having similar number of component rejections. Interestingly however, the resultant GradCAM values from the best performing networks are quite dissimilar, and pipelines with similar number of components do not necessarily result in higher CAM similarity. One possible explanation for this could be related to a common criticism of GradCam for visual recognition tasks, in which the activations often do not fully encapsulate the target object~\cite{chattopadhay2018grad}. Motor imagery is largely attributed to activity in the premotor, parietal, as well as the sensorymotor area. While activations from each pipeline denote some combinations of channels close to these cortical areas, it is still rather inconsistent, and the temporal highlight of these seem to be even more so. With scalp EEG usually the sensor measurements are already mixtures of signals from these sources, but nonetheless the variety in network activations are rather perplexing. In computer vision, neural networks have been known to implicitly learn features~\cite{olah2020naturally, schubert2021high, voss2021branch}, and often different architectures would form similar filter structures even across different tasks. Evidence from study of CNN filters also suggests filters of trained networks can serve as a proxy of data distribution, and that often similar models have little shifts in the filters across different tasks~\cite{gavrikov2022cnn}. While the network activations show a somewhat different patterns between pipelines, it is possible that the filters themselves may be more similar. Regardless, our current results are not sufficient to posit whether noise features are implicitly being ignored by the network models. For a clearer verdict on the necessity of ICA-based noise rejection methods in classification tasks, we believe further work analyzing the intermediate weight and output of the networks, as well as their connection to the output of ICA based rejection methods are necessary.

\begin{table*}
\captionsetup{justification=centering}
	\centering
	\begin{center}
    
            a)
            \begin{tabular}{|l|c|c|c|c|c|}	
				\hline	
				\textbf{SSIM-EEG}&NoICA&RunICA-MARA&RunICA-ICLabel&AMICA-MARA&AMICA-ICLabel\\
				\hline
				NoICA &1 & 0.8031 & 0.9374 & 0.7989 & 1\\ \hline
                RunICA-MARA&0.8031 & 1 & 0.7772 & 0.9695 & 0.8031\\ \hline
                RunICA-ICLabel&0.9374 & 0.7772 & 1 & 0.7716 & 0.9374\\ \hline
                AMICA-MARA&0.7989 & 0.9695 & 0.7716 & 1 & 0.7989\\ \hline
                AMICA-ICLabel&1 & 0.8031 & 0.9374 & 0.7989 & 1 \\
				\hline			
                \hline
			\end{tabular}

            b)
            \begin{tabular}{|l|c|c|c|c|c|}			
    
				\hline	
				\textbf{SSIM-CAM}&NoICA&RunICA-MARA&RunICA-ICLabel&AMICA-MARA&AMICA-ICLabel\\
				\hline
				NoICA & 1 & 0.4167 & 0.4417 & 0.4015 & 0.4036\\  \hline
                RunICA-MARA & 0.4167 & 1 & 0.4580 & 0.4660 & 0.4610\\  \hline
                RunICA-ICLabel & 0.4417 & 0.4580 & 1 & 0.4754 & 0.4655\\  \hline
                AMICA-MARA & 0.4015 & 0.4660 & 0.4754 & 1 & 0.4591\\  \hline
                AMICA-ICLabel & 0.4036 & 0.4610 & 0.4655 & 0.4591 & 1\\
				\hline			
			\end{tabular}
	\end{center}

	\caption{Structural similarity index (SSIM)~\cite{wang2004image} between pipelines of all test set EEG trials from BCIC-IV2a dataset participant 1, and of averaged GradCAM results from test set of BCIC-IV2a, participant 1. For GradCAM, the best performing model of EEGNetV4 was used to extract activation maps of the separable depth convolutional layer. Note that a SSIM of 1 does not necessarily indicate the two matrices are identical, but that they are structurally extremely similar.} 
	\label{tab:cam_ssim}	
\end{table*}

\subsection{Network performance: is it comparable to reference papers and results?}
In the interest of attempting to generalize findings across different datasets, we largely kept pre-processing pipelines before classification as consistent as possible throughout the study. Possibly as a result of this, the final prediction performance was not necessarily at state-of-the-art levels - especially for the more widely-tested BCIC-IV2a dataset. Considering that many recent SOTA studies in motor imagery classification have used task-specific feature extraction methods ~\cite{kwon2019subject, sakhavi2015parallel, sakhavi2018learning, ang2016eeg} (see ~\cite{aggarwal2019signal} for review), this is not too surprising. We, however, believe this does not invalidate our findings in that ICA pipelines have not necessarily made a considerable impact in classifier performance: taking note of the classifier performance in the reference paper of ShallowNet by Schirrmeister et al.~\cite{schirrmeister2017deep}, in which the same BCIC-IV2a dataset was trained with the same network but on the same network parameter across all participants, the overall performance difference is relatively minor: the reference paper reports an average accuracy of slightly below 70\%, while our results report an overall average test F1-score of .65. 
In the long-term memory dataset however, the final performance reported in our study was generally {\emph higher} overall than both the dataset reference paper~\cite{kang2020eeg} as well as the follow-up study using Transformer based networks~\cite{mametkulov2022explainable}. It should be noted that a lot of the specifics leading up to classification were very different, such as the use of balancing of samples across classes by undersampling the numerous class, as well as the addition of an additional, 15th participant that was not present at the time when the reference paper was published. Because replication of the dataset reference paper results nor SOTA results were not the core focus of this study, and because there are differences in implementation details among the different existing studies, it is difficult to draw a conclusion on whether the classifier was truly optimized enough. 
At the time of writing this manuscript, the original paper for the visual memory dataset is not yet available, making any comparison to reference results difficult. However, a hypothesis preprint is available at ~\cite{trubutschek2022eegmanypipelines, algermissen2021eegmanypipelines}.

Based on the breadth of pipelines as well as the size of the hyperparameter space, combined with the fact that we optimized parameters for each participant within the dataset, and the consistency of results across these networks, participants and datasets, we can only reiterate our conclusion that the benefits gained from using ICA based noise rejections are rather minimal at best for classification. We also believe the discrepancies in the implementation details warrant the need for sharing of transparent study implementation details in order to further facilitate reproduction studies to validate the findings by other scholars.

\subsection{Hyperparameter choice: How well optimized were the networks?}
In the preliminary network-non-specific parameter search portion of our study, we found clear trends for optimal values of learning rate and early stopping parameters that for the BCIC-IV2a dataset. When assessing the impact of further, dataset-specific fine-tuning of these parameters, studies have indicated that using default hyperparameter values may not adversely affect performance compared to optimized ones, especially given the trade-off between model performance and computational resources~\cite{probst2019tunability, weerts2020importance}. Hyperparameter tuning on a proxy dataset can also yield reasonable results ~\cite{shleifer2019using}, but in general the proxy datasets share more than some similar properties; as in, they are usually subsets of a larger main dataset. For a clearer conclusion on the effect of pre-processing pipelines on model performance, it may be necessary in the future to fully optimize for a larger range of hyper-parameters of the models.

In the main network-specific model parameter optimization portion of our study, it appeared that no specific parameter was preferred across the different folds and configurations. The figures in the supplementary material (Figures 7,8,9 for BCIC-IV2a dataset; Figures 10,11,12 for long-term memory dataset; Figures 13,14,15 for the Visual Memory dataset) show the optimized parameters from each participant in each pipelines among the search space, plotted against their respective final test-set F1 scores. While in some parameters (e.g. pool\_time\_length of ShallowNet, conv1\_kernel of MLSTM-FCN, and F1 and third\_kernel\_size of EEGNetv4 for the BCIC-IV2a dataset; filter\_time\_length and pool\_time\_length of ShallowNet, lstm\_num\_layers of MLSTM-FCN, F1 and D in EEGNetV4 for the long-term memory dataset; lstm\_num\_layers of MLSTM-FCN, F1 of EEGNetv4 for the visual memory dataset), there were indeed trends towards the optimal parameter, the picture is less clear in the network-specific parameter optimization. It can be seen that there were certain parameters of the same network in which the parameter space was different between the two datasets, especially for parameters related to filter sizes. This was due to the consideration of dataset time series epoch length itself; the BCIC-IV2a datset epochs span 5 seconds, while the long-term memory dataset epochs and the visual memory dataset are one second in length. This not only affects considerations for layer kernel sizes, but the characteristics of the underlying brain signals of interest as well; language long-term memory formation was best predicted on the feedback locked portion of the long-term memory dataset in which the signal of interest was within the second (partially also due to study design), whereas it is common for motor imagery signal epochs to span multiple seconds. In the context of the main goal of our study however, it appears that there are no hyperparameters in the space that are specifically favored by data from specific ICA rejection pipelines. While the actual possible search space for hyperparameters is much larger compared to the scale of our study, this finding may be relevant to our previous observation that ICA component rejections do not make notably impactful contributions to classifier performance, and therefore no preferred hyperparameter specific to specific pipelines can be found.

\subsection{Future work}
Based on our experiments, we have failed to identify a clear advantage from utilizing ICA-based artifact removal methods on EEG data in neural-network based decoding studies.
However, additional investigations are necessary to further strengthen this conclusion: while our work did attempt to make a cross-participant and dataset-representative level analysis of the effect of ICA rejection pipelines on classifier performance, for a more thorough conclusion, training and hyperparameter optimization on a cross-participant, participant-independent level is also necessary. Furthermore, in order to generalize our findings so far to the general space of brain signal classification, we believe the analysis should be expanded to cover more datasets spanning different cognitive tasks. We could possibly expand the analysis to tasks such as P300-based paradigms (e.g. Spellers), lie-detection, risk-taking, and visual/short term memory. We also believe it's worth investigating whether the lack of effect would also be observable in a simulated EEG dataset.

The choice of model architecture in our study was largely based on network models that have seen extensive use within EEG data classification tasks. Such models are overwhelmingly CNN-based architectures, although we have added an LSTM-based model to our considerations. Based on the current state-of-the-art in the literature, future work will need to extend our analyses to other model architectures, such as models based on self-supervised pre-training or transformers.
It is still unclear whether the findings can be extended to non-neural network methods, specifically conventional machine learning algorithms such as Support Vector Machines (SVM), Linear Discriminant Analysis (LDA), and decision trees. While we attempted a preliminary test with LDAs, we failed to achieve reasonable performance of the models to conduct the comparison in the first place, possibly due to a lack of explicit feature extraction methods in the interest of keeping the pre-processing consistent to our main study. For conventional machine learning methods, we believe a separate study with explicit feature extraction pipelines would be more helpful in discerning whether our observations extend to such methods as well.

\section{Acknowledgments}
This study was supported by the National Research Foundation of Korea under project BK21 FOUR and grants NRF-2022R1A2C2092118, NRF-2022R1H1A2092007, NRF-2019R1A2C2007612, as well as by Institute of Information \& Communications Technology Planning \& Evaluation (IITP) grants funded by the Korea government (No. 2017-0-00451, Development of BCI based Brain and Cognitive Computing Technology for Recognizing User’s Intentions using Deep Learning; No. RS-2019-II190079, Department of Artificial Intelligence, Korea University; No. RS-2021-II212068, Artificial Intelligence Innovation Hub). This research was also funded in part by the Austrian Science Fund (FWF) 10.55776/I6222.

\section{References}
\bibliographystyle{IEEEtran}
\bibliography{ownpubs}

\end{document}